\documentclass[11 pt,oneside,onecolumn,a4paper]{article}
\usepackage{amsmath}
\usepackage{graphicx}
\DeclareGraphicsExtensions{.eps,.ps,.pdf}
\usepackage{bbm}
\usepackage{bm}% bold math
\usepackage{epsfig}
\usepackage[T1]{fontenc}
\usepackage{esint}
\usepackage{amssymb}

\usepackage{lipsum}
\usepackage{savesym}
mm \leftmargin 5pt \rightmargin 5pt \hoffset= -15mm

\title{Regular patterns in the information flow \\ of local dephasing channels}
\author{Filippo Giraldi}

\date{\small{School of Chemistry and Physics, University of KwaZulu-Natal\\ 
and National Institute for Theoretical Physics (NITheP)\\
Westville Campus, Durban 4000, South Africa
\vspace{1em}\\Gruppo Nazionale per la
Fisica Matematica (GNFM-INdAM)\\
c/o Istituto Nazionale di Alta Matematica Francesco Severi\\
Citt\'a Universitaria, Piazza Aldo Moro 5, 00185 Roma, Italy}}

\begin{document}

\maketitle

 \begin{abstract}
Consider local dephasing processes of a qubit that interacts with a structured reservoir of frequency modes or a thermal bath, with Ohmic-like spectral density (SD). It is known that non-Markovian evolution appears uniquely above a temperature-dependent critical value of the Ohmicity parameter, and non-Markovianity can be induced by properly engineering the external environment. In the same scenario, we find that the flow of quantum information shows regular patterns: alternate directions appear in correspondence of periodical intervals of the Ohmicity parameter $\alpha_0$. The information flows back into the system over long times for $2+4n<\alpha_0<4+4n$, at zero temperature, and for $3+4n<\alpha_0<5+4n$, at non-vanishing temperatures, where $n=0,1,2,\ldots$. Otherwise, the long time information flows into the environment. In the transition from vanishing to arbitrary non-vanishing temperature, the long time back-flow of information is stable for $3+4n<\alpha_0<4+4n$, while it is reverted for $2+4n<\alpha_0<3+4n$ and $4+4n<\alpha_0<5+4n$. The patterns of the information flow are not altered if the low frequency Ohmic-like profiles of the SDs are perturbed with additional factors that consist in arbitrary powers of logarithmic forms. Consequently, the flow of information can be controlled, directed and reverted over long times by engineering a wide variety of reservoirs that includes and continuously departs from the Ohmic-like structure at low frequencies. Non-Markovianity and recoherence appear according to the same rules along with the back-flow of information. 
\end{abstract}
\vspace{-0 mm}
 \maketitle

PACS: 03.65.Yz, 03.65.Ta

\section{Introduction}\label{1}

In open quantum systems the loss, revival or maintenance of quantum correlations is deeply related to the structure of the external environment \cite{BP,Weiss}. The persistent interactions between system and environment and the appearance of memory effects were usually referred as non-Markovianity. These conditions have been largely investigated and in the last years new definitions and measures of non-Markovianity have been proposed. See Refs. \cite{BreuerNnMarkovRMP2016,PlenioNnMarkovRPP2014} for a review. Non-Markovianity can be interpreted in terms of the flow of quantum information. This quantity is defined in various ways, via the Fisher information \cite{IFlowDef1}, the
fidelity \cite{IFlowDef2} or the mutual information \cite{IFlowDef3}, to name a few. The trace-distance measure introduced in Ref. \cite{BreuernnMarkovPRL2009} estimates the relative distinguishability of two arbitrary quantum states. In Markovian processes this measure diminishes monotonically in time. This behavior can be seen as a loss of quantum information by the open system, while in non-Markovian dynamics the memory effects can be interpreted as a flow of quantum information from the external environment back into the open system. In this way, the structures of the environment that induce non-Markovian dynamics can be found by studying the direction of the flow of quantum information \cite{ManiscalcoPRAr2013,ManiscalcoNJP2015,VacchiniPRA2016,ManiscalcoPRAr2016}.

The dephasing process of qubit (two-level system) that interacts with a structured reservoir of frequency modes \cite{BP,Weiss,RH1,RH2,RH3} is a referential scenario for the study of non-Markovianity. In this system the measures of non-Markovianity mentioned above suggest the same conditions for the appearance of non-Markovian dynamics \cite{QbtMueasuresConsistentManiscalcoPRA2014} and are easily evaluated from the dephasing rate and the dephasing factor of the system \cite{nnMarkovNeg2LZouPRA2011,ManiscalcoPRAr2013,ManiscalcoNJP2015,VacchiniPRA2016,ManiscalcoPRAr2016}. In fact, persistent negative values of the dephasing rate or, equivalently, a decreasing dephasing factor indicate back-flow of information into the open system and witness non-Markovianity. 

A remarkable analysis of the dependence of non-Markovianity on the structure of the external environment is performed in Ref. \cite{ManiscalcoPRAr2013}. Convexity properties that involve the general reservoir spectrum provide conditions for the appearance of non-Markovian dynamics. 
For Ohmic-like SDs with exponential cutoff the transition from Markovian to non-Markovian dynamics appears in correspondence of a critical value of the Ohmicity parameter. In fact, temporary back-flow of information and recoherence are found uniquely for values of the Ohmicity parameters that are larger than such critical value. At zero temperature the critical value is equal to $2$, it increases monotonically with temperature and becomes $3$ at infinite temperature. See Ref. \cite{ManiscalcoPRAr2013} for details. Great efforts have been made for the experimental observation of these phenomena. Simulations of open system dynamics have been performed with trapped ions \cite{OQssTrappedIonsExp} and transition from Markovian to non-Markovian dynamics has been obtained in an all-optical experiment \cite{expNnMarkovPNP2011}, to name a few. 

As a continuation of the scenario described above, here, we consider the local dephasing process of a qubit that interacts either with a structured reservoir of frequency modes or with a thermal bath. In addition to the Ohmic-like structure, the SDs under study include at low frequencies removable logarithmic singularities \cite{GiraldiLogSD,GiraldiEPJD2015}. 
At higher frequencies the SDs are arbitrarily shaped. We study the decoherence and recoherence processes by evaluating the dephasing factor and we investigate the flow of quantum information by analyzing the dephasing rate, in dependence on the structure of the reservoir or of the thermal bath. We also search for regular patterns in the direction of the flow of information that allow a full manipulation of the flow itself and, consequently, of non-Markovianity and recoherence by engineering the low frequency structure of the external environment.

The paper is organized as follows. Section \ref{2} is devoted
to the description of the model. In Section \ref{3}, asymptotic coherence is described in term of integral properties of the SD. Section \ref{31} is devoted to the description of the Ohmic-like spectral densities with additional removable logarithmic singularities. The decoherence and recoherence processes are studied in Section \ref{32}, by analyzing the dephasing factor both at zero and non-vanishing temperature. Patterns in the flow of quantum information are shown in Section \ref{4}. Conclusions are drawn in Section \ref{5}, and details on the calculations are provided in the Appendix.

\section{The model}\label{2}

The system of a qubit that interacts locally with a reservoir of frequency modes is described by the following microscopic Hamiltonian \cite{BP,Weiss,ManiscalcoPRAr2013,ManiscalcoNJP2015},
\begin{equation}
H= \omega_0 \sigma_z+ \sum_k \omega_k b^{\dagger}_k b_k
+\sum_k \sigma_z \left(g_k a_k+g^{\ast}_k a^{\dagger}_k\right), \label{H}
\end{equation}
in units where $\hbar=1$.
 The transition frequency of the qubit is $\omega_0$, while 
$\sigma_z$ represents the $z$-component Pauli spin operator. The index $k$ runs over the frequency modes. The parameter $\omega_k$ represents the frequency of the $k$th mode, while $b^{\dagger}_k$ and $b_k$ are the rising and lowering operator, respectively, of the same mode.
The coefficient $g_k$ represents the coupling strength between the qubit and the $k$th frequency mode. 
The reduced density matrix $\rho(t)$ represents the mixed state of the qubit at the time $t$ and is obtained by tracing the density matrix of the whole system at the time $t$ over the Hilbert space of the external environment \cite{BP}. The model is exactly solvable \cite{RH1,RH2,RH3}. 

Let the qubit be initially decoupled from the external environment that is represented by a structured reservoir of field modes or by a thermal bath. The reduced time evolution is described in the interaction picture by the following master equation,
\begin{equation}
\dot{\rho}(t)=\gamma(t)\left(\sigma_z \rho(t) \sigma_z -\rho(t)\right).  \label{Eq1}
\end{equation}
The function $\gamma(t)$ represents the dephasing rate and is related to the temperature of the thermal bath. At zero temperature, $T=0$, the dephasing rate is labeled here as $\gamma_0(t)$ and reads
\begin{eqnarray}
&&\gamma_0(t)=\int_0^{\infty} \frac{J\left(\omega\right)}{\omega}\,
\sin \left(\omega t \right) \, d\omega.  \label{gamma0} 
\end{eqnarray}
The function $J\left(\omega\right)$ represents the SD of the system and is defined in terms of the coupling constants $g_k$ via the following form,
$J\left(\omega\right)=\sum_k \left|g_k\right|^2 \delta \left(\omega-\omega_k\right)$. 
If the external environment is initially in a thermal state, $T>0$, the dephasing rate is represented here as $\gamma_T(t)$ and reads
\begin{equation}
\gamma_T(t)=\int_0^{\infty} \frac{J\left(\omega\right)}{\omega}\,
 \coth \frac{\hbar \omega}{2 k_B T}\,\sin \left(\omega t \right) d\omega.  \label{gammaT}
\end{equation}
The parameter $k_B$ is the Boltzmann constant.

The quantum coherence between the states $|0\rangle$ and $|1\rangle$ of the qubit is described by the off-diagonal element $\rho_{0,1}(t)$ of the density matrix that undergoes the following evolution \cite{RH1,RH2,RH3}, 
\begin{equation}
\rho_{0,1}(t)=\rho^{\ast}_{1,0}(t)=\rho_{0,1}(0)\, \exp \left\{-\Xi(t)\right\}.  \label{rho01t}
\end{equation}
The function $\Xi(t)$ represents the dephasing factor and depends on the temperature $T$ of the thermal bath and on the coupling between the system and the environment. At zero temperature, $T=0$, the dephasing factor is indicated here as $\Xi_0(t)$ and results in the following form,
\begin{eqnarray}
&&\Xi_0(t)=\int_0^{\infty} J\left(\omega\right)\, \frac{1- \cos\left(\omega t \right) }{\omega^2}\, d\omega.  \label{Xi0t}
\end{eqnarray}
If the external environment is initially in a thermal state, $T>0$, the dephasing factor is represented here as $\Xi_T(t)$ and reads
\begin{eqnarray}
\Xi_T(t)=\int_0^{\infty}\frac{J\left(\omega\right)}{\omega^2} \left(1-\cos\left(\omega t\right)\right)\coth \frac{\hbar \omega}{2 k_B T}. \label{XiTt}
\end{eqnarray}
 Both for vanishing and non-vanishing temperature, the dephasing factor is related to the dephasing rate via the time derivative, $\gamma_0(t)=\dot{\Xi}_0(t)$ and $\gamma_T(t)=\dot{\Xi}_T(t)$.
According to Eq. (\ref{rho01t}), recoherence corresponds to negative values of the dephasing rate.

\section{Coherence}\label{3}

The loss or persistence of coherence between the two energy eigenstates of the qubit depends on integral and low frequency properties of the SDs.
At zero temperature, $T=0$, coherence is not entirely lost over long times if the second negative moment of the SD is finite,
\begin{eqnarray}
\int_0^{\infty}\frac{ J\left(\omega\right)}{\omega^2}\,
d \omega<\infty. \label{cond1} 
\end{eqnarray}
Under this condition the coherence term shows persistence of residual coherence,
\begin{eqnarray}
&&\hspace{-2em}\rho_{0,1}\left(\infty\right)=\rho_{0,1}(0)\exp \left\{- 
\int_0^{\infty}\frac{ J\left(\omega\right)}{\omega^2}\,
d \omega \right\}. \label{rhoinfty}
\end{eqnarray}
If the external environment consists in a thermal bath, $T>0$, and the following condition holds,
\begin{eqnarray}
\int_0^{\infty}
\frac{ J\left(\omega\right)}{\omega^2}\,\coth \frac{\hbar \omega}{2 k_B T}\,d \omega<\infty, \label{cond1T}
\end{eqnarray}
 the coherence term tends over long times to the non-vanishing asymptotic value
\begin{eqnarray}
&&\hspace{-2em}\rho_{0,1}\left(\infty\right)=\rho_{0,1}(0)\exp \Bigg\{- 
\int_0^{\infty}
\frac{ J\left(\omega\right)}{\omega^2}%\nonumber \\&&\hspace{2.9em}\times \,
\coth \frac{\hbar \omega}{2 k_B T} \,d \omega\Bigg\}.\label{rhoinftyT}
\end{eqnarray}
The maximum modulus of the ratio between asymptotic and initial coherence is obtained at zero temperature, $T=0$, from Eq. (\ref{rhoinfty}).

According to Eq. (\ref{rho01t}), residual coherence persists over long times if the dephasing factor does not diverge asymptotically, while coherence is fully lost if the dephasing factor diverges. Consequently, the dependence of coherence on the structure of the SD can be analyzed via the dephasing factor itself. We intend to study the short and long time behavior of the dephasing factor for a large variety of SDs. We consider low frequency structures of the SDs such that, for $T=0$, the second negative moment is either finite, Eq. (\ref{cond1}), or infinite,
\begin{eqnarray}
\int_0^{\infty}\frac{ J\left(\omega\right)}{\omega^2}\,
d \omega=\infty.\label{cond2}
\end{eqnarray}
At non-vanishing temperatures, $T>0$, we choose low frequency profiles of the SDs that fulfill either the constraint (\ref{cond1T}) or the following one,
\begin{eqnarray}
\int_0^{\infty}
\frac{ J\left(\omega\right)}{\omega^2}\,\coth \frac{\hbar \omega}{2 k_B T}\,d \omega=\infty. \label{cond2T}
\end{eqnarray}
Details on the structure of the SDs under study are provided below.

\section{Spectral densities with removable logarithmic singularities}\label{31}

The fast development of quantum technologies allows the engineering of the most various environments. According to the remarkable analysis performed in Refs. \cite{SDeng1,SDengBECM2011}, an impurity that is trapped in a double-well potential and is surrounded by a cold gas reproduces, under suitable conditions, a qubit that interacts with an Ohmic-like environment. The Ohmicity parameter increases by enhancing the scattering length that is related to the boson-boson coupling \cite{SDengBECM2011}. In case the gas is free and one-dimensional, the SD changes from sub-Ohmic to Ohmic and to super-Ohmic by increasing the scattering length. In the two-dimensional non-interacting condition the spectrum is Ohmic and the super-Ohmic regime is obtained if the magnitude of the interaction decreases. The SD is super-Ohmic in the non-interacting condition if the gas is three-dimensional. We refer to \cite{SDengBECM2011} for details.

In light of the above observation we focus on SDs that are Ohmic-like at low frequencies and are arbitrarily shaped at higher frequency. We intend to analyze the feature of the open dynamics, the flow of quantum information, non-Markovianity and recoherence of the qubit. We also evaluate the accuracy of the results obtained for the experimentally feasible Ohmic-like SDs by perturbing the power laws of the Ohmic-like profiles with additional factors that are represented by arbitrary powers of logarithmic forms. In this way, we consider a wide variety of SDs that includes and continuously departs from the Ohmic-like condition. Positive (negative) or vanishing values of the first logarithmic power enhance (reduce) or unchange the power law profile, and, especially, provide legitimate SDs \cite{GiraldiLogSD}.

For the sake of convenience, the SDs $J \left( \omega\right)$ are described via the dimensionless auxiliary function $\Omega\left(\nu\right)$. This function is defined for every $\nu\geq 0$ by the following scaling property, $J \left(   \nu \Delta\right)/\Delta=  \Omega\left(\nu \right)$, in terms of a general scale frequency $\Delta$ of the system. At non-vanishing temperatures it is convenient to define the effective SD $J_T\left(\omega\right)$ as follows, 
\begin{eqnarray}
&&\hspace{-0em}J_T\left(\omega\right)=J\left(\omega\right) \coth \frac{\hbar \omega}{2 k_B T}. \label{JT}
\end{eqnarray}
The corresponding auxiliary function $\Omega_T\left(\nu\right)$ 
reads $\Omega_T\left(\nu\right)= \Omega\left(\nu\right)
\coth \left(\left(\hbar \Delta \nu\right)/\left(2 k_B T\right)\right)$. In this way, the action of the thermal bath can be represented via a transformed SD. The first class of SDs under study is defined by auxiliary functions $\Omega\left(\nu\right)$ that are continuous for every $\nu> 0$ and exhibit the following asymptotic behavior \cite{BleisteinBook} as $\nu\to 0^+$,
\begin{eqnarray}
&&\hspace{-4em}\Omega\left(\nu\right)\sim 
\sum_{j=0}^{\infty}
\sum_{k=0}^{n_j}c_{j,k} \nu^{\alpha_j}\left(- \ln \nu\right)^k,  \label{o0log} 
\end{eqnarray}
 where $\xi>0$, 
$\infty> n_j\geq0$, $\alpha_{j+1}>\alpha_j$, for every $j\geq 0$, and $\alpha_j\uparrow +\infty$ as $j\to +\infty$. Furthermore, we consider $\alpha_0\geq 0$, and $n_0=0$ if $\alpha_0=0$. The power $\alpha_0$ is referred as the Ohmicity parameter \cite{BP,Weiss}. In fact, if $n_0=0$, the corresponding SDs are super-Ohmic for $\alpha_0>1$, Ohmic for $\alpha_0=1$ and sub-Ohmic for $1>\alpha_0>0$, as $\omega\to 0^+$. The singularity in $\nu=0$ is removable by defining $\Omega(0)$ as the finite limit of $\Omega\left(\nu\right)$ as $\nu\to0^+$. Notice that Eq. (\ref{o0log}) describes a large variety of asymptotic forms that include exponential and stretched exponential functions, inverse power laws and natural powers of logarithmic forms. The summability of the SD is guaranteed by the constraint $\Omega\left(\nu\right)= O\left(\nu^{-1-\chi_0}\right)$, as $\nu\to+\infty$, where $\chi_0>0$. Additionally, the Mellin transforms $\hat{\Omega}\left(s\right)$ and $\hat{\Omega}_T\left(s\right)$ of the auxiliary functions $\Omega\left(\nu\right)$ and $\Omega_T\left(\nu\right)$, and the meromorphic continuations \cite{BleisteinBook,Wong-BOOK1989} are required to decay sufficiently fast as $\left|\operatorname{Im}s \right|\to+\infty$. See Appendix for details.

In light of the asymptotic analysis performed in Refs. \cite{WangLinJMAA1978,Wong-BOOK1989}, the second class of SDs under study is described by auxiliary functions with the following asymptotic expansion as $\nu\to 0^+$, 
\begin{eqnarray}
&&\hspace{-4em}\Omega\left(\nu\right)\sim \sum_{j=0}^{\infty}w_j
\, \nu^{\alpha_j} \left(-\ln \nu\right)^{\beta_j}.  \label{OmegaLog0}
\end{eqnarray}
The powers $\beta_j$ are real valued, while $\alpha_0>0$. The logarithmic singularity in $\nu=0$ is removed by setting $\Omega(0)=0$. Let the parameter $\bar{n}$ be the least natural number such that $\alpha_{k-1}+1\leq \bar{n}<\alpha_{k}+1$, where the index $k$ is a non-vanishing natural number. 
The function $\Omega^{\left(\bar{n}\right)}\left(\nu\right)$ is required to be continuous on the interval $\left(0,\infty\right)$. The integral $\int_0^{\infty}\Omega\left(\nu\right)\exp\left\{-\imath \xi \nu\right\} d \nu$ must converge uniformly for all sufficiently large values of the variable $\xi$ and the integral $\int \Omega^{\left(\bar{n} \right)}\left(\nu\right)\exp\left\{-\imath \xi \nu\right\} d \nu$ has to converge at $\nu=+\infty$ uniformly for all sufficiently large values of the variable $\xi$. The auxiliary function is required to be differentiable $k$ times and the following asymptotic expansion at $\nu\to 0^+$,
$$\Omega^{(k)}\left(\nu\right)\sim \sum_{j=0}^{\infty}w_j
\, \frac{d^k}{d\nu^k}\left(\nu^{\alpha_j} \left(-\ln \nu\right)^{\beta_j}\right),$$
is required to hold for every $k=0,1, \ldots,\bar{n} $. Furthermore, for every $k=0, \ldots,\bar{n}-1$, the function $\Omega^{(k)}\left(\nu\right)$ has to vanish as $\nu\to +\infty$.

If compared to the first class, the second class of SDs has to fulfill more constraints but includes arbitrary powers of logarithmic forms. 
In both the classes under study the auxiliary functions $\Omega\left(\nu\right)$ are non-negative, bounded and summable, due to physical grounds, and, apart from the above constraints, arbitrarily shaped at high frequencies \cite{GiraldiLogSD}.

\section{The dephasing factor}\label{32}

We start the analysis of the dephasing factor by considering a reservoir of frequency modes at zero temperature, $T=0$, as external environment. The SDs under study belong to the first class introduced in Section \ref{31}. Over short times, $t \ll 1/\Delta$, the dephasing factor increases quadratically in time,	
\begin{eqnarray}
&&\hspace{-2em}\Xi_0(t)\sim \frac{l_0}{2}\, t^2,
 \label{Xi000}
\end{eqnarray}
where $l_0=\int_0^{\infty} J \left(\omega\right) d\omega$.
This behavior is independent of the low or high frequency structure of the SDs under study. Instead, the evolution of the dephasing factor over long times, $t \gg 1/ 
\Delta$, is various and is determined by the low frequency structure of the SD, given by Eq. (\ref{o0log}).
If $\alpha_0=n_0=0$ the dephasing factor grows linearly for $t\gg 1/\Delta$,
\begin{eqnarray}
\hspace{-0.9em}\Xi_0(t)
\sim \frac{\pi c_{0,0}}{2} \,\Delta t. 
 \label{Xi00}
\end{eqnarray}
If $0< \alpha_0<1$ we find for $t \gg 1/\Delta$ the following divergent behavior,
\begin{eqnarray}
\hspace{-0.9em}\Xi_0(t)\sim c_{0,n_0} r_1 \,\left(\Delta t\right)^{1-\alpha_0}
\ln^{n_0}\left(\Delta t\right), 
 \label{Xi01}
\end{eqnarray}
where $r_1= \sin\left(\pi \alpha_0/2\right)\Gamma\left(\alpha_0\right)/\left(1-\alpha_0\right)$. The above form grows as a power law for $n_0=0$,
\begin{eqnarray}
\hspace{-0.9em}\Xi_0(t)\sim c_{0,n_0} r_1 \,\left(\Delta t\right)^{1-\alpha_0}.
 \label{Xi01pl}
\end{eqnarray}
 If $\alpha_0=1$ we obtain over long times, $t \gg 1/\Delta$, the divergent logarithmic form as below,
\begin{eqnarray}
\hspace{-0.9em}\Xi_0(t)\sim \frac{c_{0,n_0}}{n_0+1} 
\ln^{n_0+1}\left(\Delta t\right). 
\label{Xi02}
\end{eqnarray}
If $\alpha_0>1$ and $\alpha_0$ is not an even natural number, the dephasing factor tends to the following asymptotic value,
\begin{eqnarray}
\Xi_0\left(\infty\right)=\int_0^{\infty}\frac{J\left(\omega\right)}{\omega^2}\, d \omega, 
 \label{Xi03}
\end{eqnarray}
according to relaxations that involve logarithmic forms,
\begin{eqnarray}
\hspace{-0.9em}\Xi_0(t)\sim \Xi_0\left(\infty\right)+ c_{0,n_0} r_1 \,\left(\Delta t\right)^{1-\alpha_0}
\ln^{n_0}\left(\Delta t\right). 
 \label{Xi04}
\end{eqnarray}
The above expression turns into inverse power laws for $n_0=0$,
\begin{eqnarray}
\hspace{-0.9em}\Xi_0(t)\sim \Xi_0\left(\infty\right)+ c_{0,n_0} r_1 \,\left(\Delta t\right)^{1-\alpha_0}. 
 \label{Xi04pl}
\end{eqnarray}
If $\alpha_0>1$ and $\alpha_0=2m_0$, where $m_0$ and $n_0$ are non-vanishing natural numbers, we find
\begin{eqnarray}
\hspace{-0.9em}\Xi_0(t)\sim \Xi_0\left(\infty\right)+
 c_{0,n_0} r^{\prime}_1 \,\left(\Delta t\right)^{1-2m_0}
\ln^{n_0-1}\left(\Delta t\right), 
 \label{Xi0even}
\end{eqnarray}
where
$r^{\prime}_1=\pi (-1)^{m_0}n_0 \left(2 m_0-2\right)!/2$.
The above relaxation provides power laws for $n_0=1$,
\begin{eqnarray}
\hspace{-0.9em}\Xi_0(t)\sim \Xi_0\left(\infty\right)+
 c_{0,n_0} r^{\prime}_1 \,\left(\Delta t\right)^{1-2m_0}. 
 \label{Xi0evenPL}
\end{eqnarray}
If $\alpha_0$ is an even natural number and $n_0$ vanishes, consider the least non-vanishing index $k_0$ such that either $\alpha_{k_0}$ does not take even natural values or $\alpha_{k_0}=2m_{k_0}$, where the natural numbers $m_{k_0}$ and $n_{k_0}$ do not vanishes. The function $\Xi_{0}(t)$ is obtained in the former case from Eqs. (\ref{Xi04}) and (\ref{Xi04pl}) by substituting the power $\alpha_0$ with $\alpha_{k_0}$ and $n_0$ with $n_{k_0}$, or in the latter case from Eqs. (\ref{Xi0even}) and (\ref{Xi0evenPL}) by substituting the power $m_0$ with $m_{k_0}$ and $n_0$ with $n_{k_0}$. We consider SDs such that the index $k_0$ exists.

At this stage we focus on SDs that belong to the second class introduced in Section \ref{31} with a finite negative second moment, Eq. (\ref{cond1}). This conditions requires that the Ohmicity parameter $\alpha_0$ is larger than unity, $\alpha_0>1$. Over long times, $t \gg 1/\Delta$, the dephasing factor relaxes to the asymptotic value $\Xi_0 \left(\infty\right)$ according to arbitrarily positive or negative, or vanishing powers of logarithmic forms, 
\begin{eqnarray}
&&\Xi_0(t)\sim 
\Xi_0\left(\infty\right)+w_0\left(\Delta t\right)^{1-\alpha_0}\Big( r_1 \ln^{\beta_0}\left(\Delta t\right)%\nonumber \\&&\hspace{3.6em}
+ \bar{r}_1\,
\ln^{\beta_0-1}\left(\Delta t\right)\Big), 
 \label{Xi0J2}
\end{eqnarray}
where $\bar{r}_1=\beta_0 \sin\left(\pi \alpha_0/2\right)
\left(\Gamma^{(1)}\left(\alpha_0-1\right)+\pi \Gamma\left(\alpha_0-1\right)/2\right)$. If the Ohmicity parameter $\alpha_0$ is not an even natural number, the dominant part of the above relaxation is equivalent to the following one, $\Xi_0(t)\sim 
\Xi_0\left(\infty\right)+w_0 r_1\left(\Delta t\right)^{1-\alpha_0}  \ln^{\beta_0}\left(\Delta t\right)$, and provides the inverse power laws $\Xi_0(t)\sim 
\Xi_0\left(\infty\right)+w_0\left(\Delta t\right)
^{1-\alpha_0}$ if $\beta_0=0$. If the Ohmicity parameter $\alpha_0$ takes even natural values, Eq. (\ref{Xi0J2}) gives $\Xi_0(t)\sim 
\Xi_0\left(\infty\right)+w_0\bar{r}_1\left(\Delta t\right)^{1-\alpha_0} 
\ln^{\beta_0-1}\left(\Delta t\right)$, and turns into the 
inverse power law relaxations $\Xi_0(t)\sim 
\Xi_0\left(\infty\right)+w_0\bar{r}_1\left(\Delta t\right)^{1-\alpha_0}$ if $\beta_0=1$.
\begin{figure}[t]
\centering
\includegraphics[height=6.25 cm, width=10.25 cm]{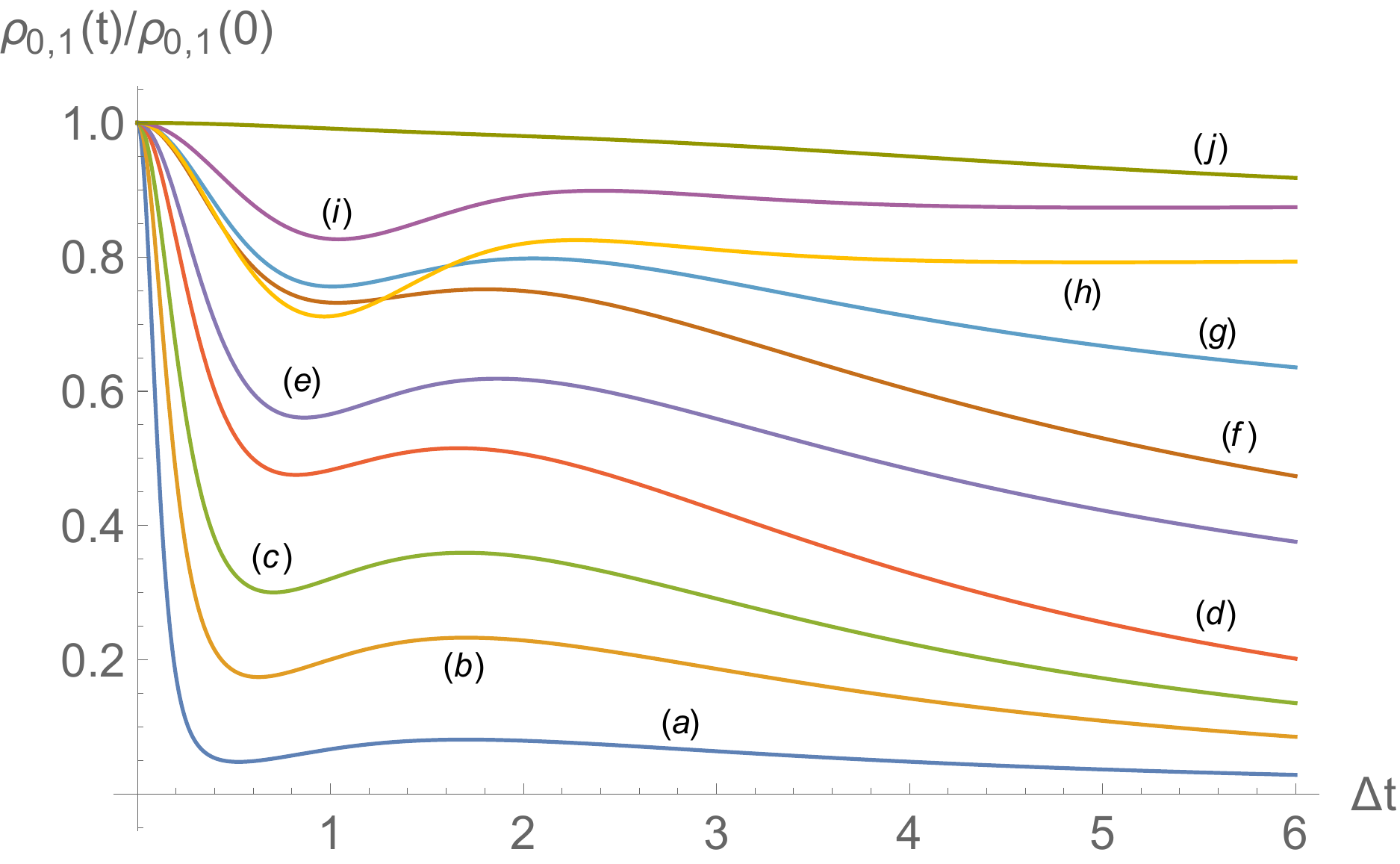}
\vspace{0.2cm}
\caption{(Color online) The quantity 
$ \rho_{0,1}(t)/\rho_{0,1}(0)$ versus $\Delta t$ for $0\leq\Delta t\leq 6$, $J\left(\omega\right)=\Delta \left(\omega/\Delta\right)^{\alpha}\exp\left\{- \lambda \omega/\Delta \right\}\ln^2\left(\omega/\Delta \right)$ and different values of the parameters $\alpha$ and $\lambda$. The curve $(a)$ corresponds to $\alpha=1.6$, $\lambda=0.3$; $(b)$ corresponds to $\alpha=1.6$, $\lambda=0.4$, $(c)$ corresponds to $\alpha=1.6$, $\lambda=0.48$; $(d)$ corresponds to $\alpha=1.6$, $\lambda=0.6$; $(e)$ corresponds to $\alpha=2$, $\lambda=0.8$; $(f)$ corresponds to $\alpha=2$, $\lambda=1$; $(g)$ corresponds to $\alpha=2.5$, $\lambda=1.2$; $(h)$ corresponds to $\alpha=5$, $\lambda=2$; $(i)$ corresponds to $\alpha=5$, $\lambda=2.2$, $(j)$ corresponds to $\alpha=3$, $\lambda=3$. Over long times the curves tend to non-vanishing values.}
\label{Fig1}
\end{figure}

\begin{figure}[t]
\centering
\includegraphics[height=6.25 cm, width=10.25 cm]{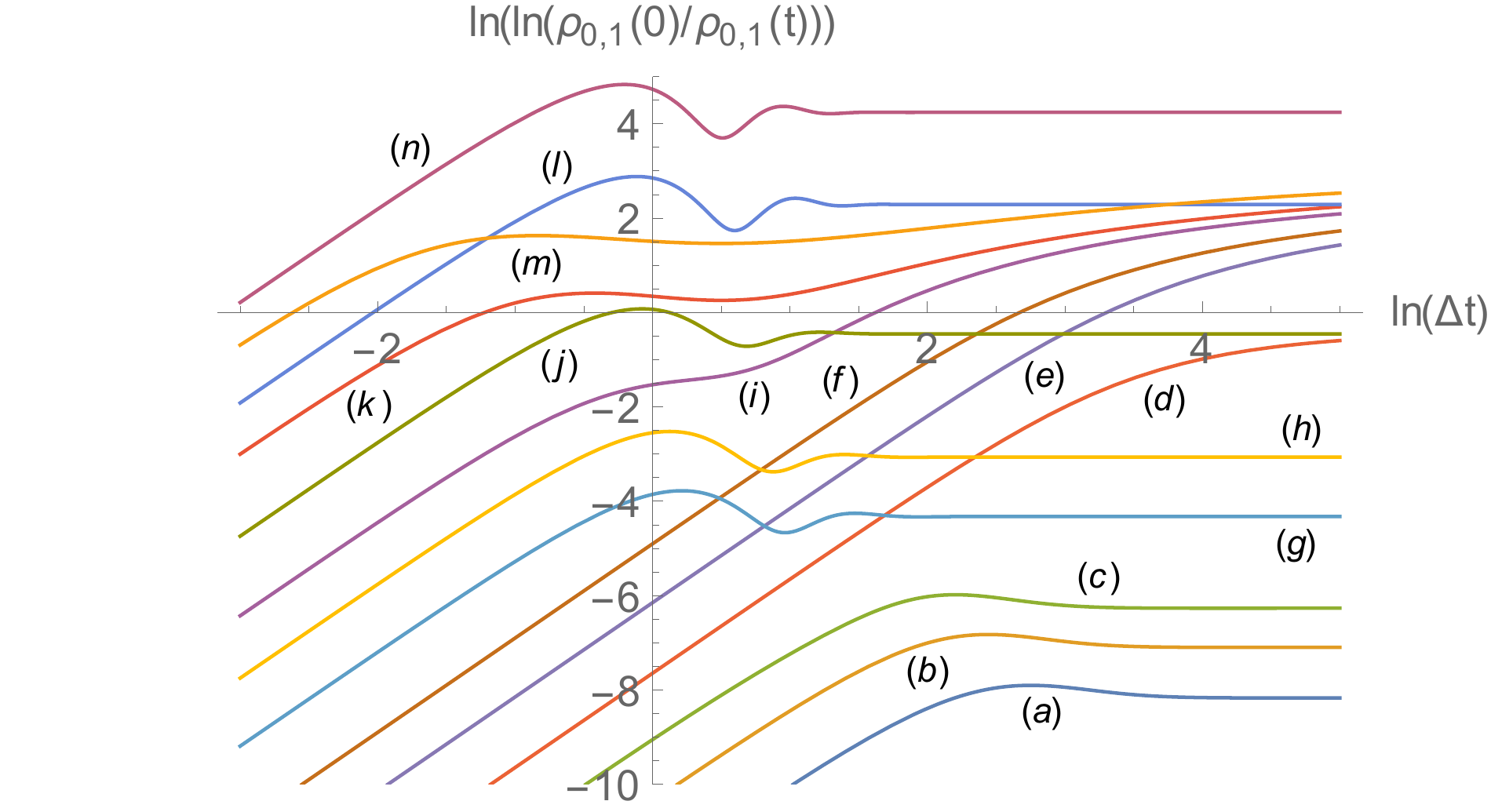}
\vspace*{0cm}
\caption{(Color online) The quantity 
$ \ln\left(\ln \left(\rho_{0,1}(0)/
\rho_{0,1}(t)\right)\right)$ versus $\ln \left(\Delta t\right)$ for $\exp\left\{-3\right\}\leq\Delta t\leq \exp\left\{5\right\}$, $J\left(\omega\right)=\Delta \left(\omega/\Delta\right)^{\alpha}\exp\left\{- \lambda \omega/\Delta \right\}\ln^2\left(\omega/\Delta \right)$ and different values of the parameters $\alpha$ and $\lambda$. The curve $(a)$ corresponds to $\alpha=5$, $\lambda=15$; $(b)$ corresponds to $\alpha=5$, $\lambda=10$, $(c)$ corresponds to $\alpha=5$, $\lambda=7$; $(d)$ corresponds to $\alpha=2$, $\lambda=22$; $(e)$ corresponds to $\alpha=1.5$, $\lambda=20$; $(f)$ corresponds to $\alpha=1.5$, $\lambda=9$; $(g)$ corresponds to $\alpha=10$, $\lambda=4.8$; $(h)$ corresponds to $\alpha=10$, $\lambda=4.3$; $(i)$ corresponds to $\alpha=1.5$, $\lambda=1$, $(j)$ corresponds to $\alpha=10$, $\lambda=3.4$; $(k)$ corresponds to $\alpha=1.5$, $\lambda=0.4$; $(l)$ corresponds to $\alpha=20$, $\lambda=6.1$; $(m)$ corresponds to $\alpha=1.5$, $\lambda=0.2$; $(n)$ corresponds to $\alpha=20$, $\lambda=5.55$. Over short times each curve tends to an asymptotic line with the slope $2$.}
\label{Fig2}
\end{figure}

\begin{figure}[t]
\centering
\includegraphics[height=6.25 cm, width=10.25 cm]{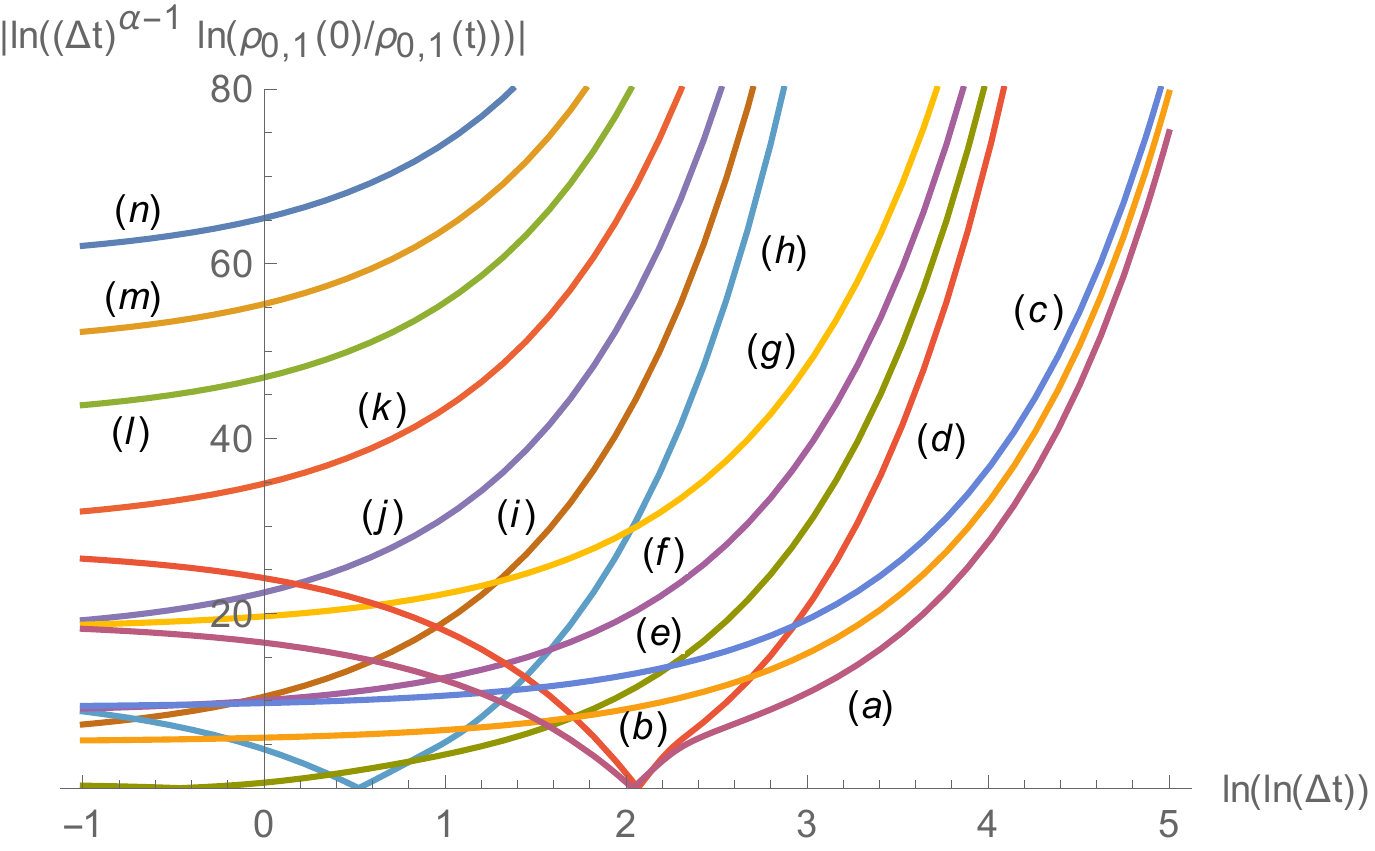}
\vspace{0.2cm}
\caption{(Color online) The quantity 
$ \left|\ln\left(\left(\Delta t\right)^{\alpha-1}\ln \left(\rho_{0,1}(0)/
\rho_{0,1}(t)\right)\right)\right|$ versus $\ln \left(\ln \left(\Delta t\right)\right)$ for $\exp\left\{\exp\left\{-1\right\}\right\}\leq\Delta t\leq \exp\left\{\exp\left\{5\right\}\right\}$, $J\left(\omega\right)=\Delta \left(\omega/\Delta\right)^{\alpha}\exp\left\{- \lambda \omega/\Delta \right\}\ln^2\left(\omega/\Delta \right)$ and different values of the parameters $\alpha$ and $\lambda$. The curve $(a)$ corresponds to $\alpha=1.5$, $\lambda=10000$; $(b)$ corresponds to $\alpha=1.5$, $\lambda=0.01$, $(c)$ corresponds to $\alpha=1.5$, $\lambda=0.0001$; $(d)$ corresponds to $\alpha=2.5$, $\lambda=10000$; $(e)$ corresponds to $\alpha=2.5$, $\lambda=1$; $(f)$ corresponds to $\alpha=2.5$, $\lambda=0.02$; $(g)$ corresponds to $\alpha=2.5$, $\lambda=0.0001$; $(h)$ corresponds to $\alpha=6$, $\lambda=10$; $(i)$ corresponds to $\alpha=6$, $\lambda=0.8$, $(j)$ corresponds to $\alpha=6$, $\lambda=0.1$; $(k)$ corresponds to $\alpha=6$, $\lambda=0.01$; $(l)$ corresponds to $\alpha=6$, $\lambda=0.001$; $(m)$ corresponds to $\alpha=6$, $\lambda=0.0002$; $(n)$ corresponds to $\alpha=6$, $\lambda=0.00003$. Over long times each curve tends to an asymptotic line and the the slope depends on the value of the parameter $\alpha$.}
\label{Fig3}
\end{figure}

\subsection{Thermal bath}

Let the external environment consist in a thermal bath, $T>0$. If the SD belongs to the first class under study with $\alpha_0>0$ the dephasing factor results again in a quadratic function of time for $t \ll 1/\Delta$,
\begin{eqnarray}
\hspace{-0.9em}\Xi_T(t)\sim \frac{l_T}{2}\,t^2,
 \label{Xi000T}
\end{eqnarray}
where $l_T=\int_0^{\infty} J \left(\omega\right) \coth \left(\hbar \omega/\left(2 k_B T \right)\right) d\omega$.
This behavior is again independent of the low or high frequency structure of the SD. Instead, the evolution of the dephasing factor over long times, $t \gg 1/ \Delta$, exhibits various behaviors in dependence on the low frequency structure 
of the SD. If $0< \alpha_0<2$ and $\alpha_0\neq1$ the dephasing factor diverges for $t \gg 1/\Delta$ as below,
\begin{eqnarray}
\hspace{-0.9em}\Xi_T(t)\sim c_{0,n_0} r_T \,\left(\Delta t\right)^{2-\alpha_0}
\ln^{n_0}\left(\Delta t\right), 
 \label{Xi00T}
\end{eqnarray}
where $r_T=2 k_B T  \cos\left(\pi\alpha_0/2\right)\Gamma\left(\alpha_0-2\right) /\left(\hbar \Delta\right) $. 
Again, power laws appear from the above conditions for $n_0=0$,
\begin{eqnarray}
\hspace{-0.9em}\Xi_T(t)\sim c_{0,n_0} r_T \,\left(\Delta t\right)^{2-\alpha_0}.
 \label{Xi00Tpl}
\end{eqnarray}
 If $\alpha_0=1$ the dephasing factor diverges for $t \gg 1/\Delta$ as below,
\begin{eqnarray}
\hspace{-0.9em}\Xi_T(t)\sim c_{0,n_0} r_T^{\prime} \,\left(\Delta t\right)
\ln^{n_0}\left(\Delta t\right), 
 \label{Xi001T}
\end{eqnarray}
where $r_T^{\prime}=\pi\,k_B T/\left(\hbar \Delta\right)$. 
The divergence becomes linear in time for $\alpha_0=1$ and $n_0=0$,
\begin{eqnarray}
\hspace{-0.9em}\Xi_T(t)\sim c_{0,n_0} r_T^{\prime} \,\left(\Delta t\right).
 \label{Xi001Tlin}
\end{eqnarray}
If $\alpha_0=2$ the dephasing factor grows for $t \gg 1/\Delta$ according to natural powers of logarithmic forms,
\begin{eqnarray}
\hspace{-0.9em}\Xi_T(t)\sim c_{0,n_0} r_T^{\prime \prime} \,
\ln^{n_0+1}\left(\Delta t\right), 
 \label{Xi002T}
\end{eqnarray}
where $r_T^{\prime \prime}=2k_B T/\left(\hbar \Delta\left(n_0+1\right)\right)$. If $\alpha_0 >2$ and $\alpha_0$ is not an odd number, the dephasing factor converges for $t \gg 1/\Delta$ to the asymptotic value
\begin{eqnarray}
\hspace{-0.9em}\Xi_T\left(\infty\right)=\int_0^{\infty}\frac{ J\left(\omega\right)}{\omega^2}\,\coth \frac{\hbar \omega}{2 k_B T }
d \omega,
 \label{Xi003T}
\end{eqnarray}
according to the following relaxations,
\begin{eqnarray}
\hspace{-0.9em}\Xi_T(t)\sim \Xi_T\left(\infty\right)+ c_{0,n_0} r_T \,\left(\Delta t\right)^{2-\alpha_0}
\ln^{n_0}\left(\Delta t\right),
 \label{Xi004T}
\end{eqnarray}
that become inverse power laws for $n_0=0$,
\begin{eqnarray}
\hspace{-0.9em}\Xi_T(t)\sim \Xi_T\left(\infty\right)+ c_{0,n_0} r_T \,\left(\Delta t\right)^{2-\alpha_0}. 
 \label{Xi004Tpl}
\end{eqnarray}
If $\alpha_0=2m_1+1$, where $m_1$ is a natural number and $n_0$ does not vanish, we find
\begin{eqnarray}
\hspace{-0.9em}\Xi_T(t)\sim \Xi_T\left(\infty\right)+
 c_{0,n_0} r^{\prime \prime \prime}_T \,\left(\Delta t\right)^{1-2m_1}
\ln^{n_0-1}\left(\Delta t\right), 
 \label{XiTodd}
\end{eqnarray}
where
$r^{\prime \prime \prime}_T=\pi (-1)^{m_1} k_B T 
n_0 \left(2 m_1-2\right)!/\left(\hbar \Delta\right)$.
The above relaxation becomes a power law for $n_0=1$,
\begin{eqnarray}
\hspace{-0.9em}\Xi_T(t)\sim \Xi_T\left(\infty\right)+
 c_{0,n_0}  r^{\prime \prime \prime}_T \,\left(\Delta t\right)^{1-2m_1}. 
 \label{XiToddPL}
\end{eqnarray}
If $\alpha_0$ is an odd natural number and $n_0$ vanishes, consider the least non-vanishing index $k_1$ such that either $\alpha_{k_1}$ does not take odd natural values or $\alpha_{k_1}=2m_{k_1}+1$, where the natural numbers $m_{k_1}$ and $n_{k_1}$ do not vanishes. The function $\Xi_{T}(t)$ is obtained in the former case from Eqs. (\ref{Xi004T}) and (\ref{Xi004Tpl}) by substituting the power $\alpha_0$ with $\alpha_{k_1}$ and $n_0$ with $n_{k_1}$, or in the latter case from Eqs. (\ref{XiTodd}) and (\ref{XiToddPL}) by substituting the power $m_1$ with $m_{k_1}$ and $n_0$ with $n_{k_1}$. We consider SDs such that the index $k_1$ exists.

At this stage we focus on SDs such that the auxiliary functions $\Omega_T\left(\nu\right)$ belong to the second class under study and that exhibit a finite second negative moment, Eq. (\ref{cond1T}). This constraint requires the Ohmicity parameter $\alpha_0$ to be larger than $2$. Under this condition, $\alpha_0>2$, a variety of logarithmic relaxations of the dephasing factor to the asymptotic value $\Xi_T\left(\infty\right)$ are obtained for $t \gg 1/\Delta$,
\begin{eqnarray}
&&\hspace{-1em}\Xi_T(t)\sim 
\Xi_T\left(\infty\right)+ w_0 \left(\Delta t\right)^{2-\alpha_0}\Big(r_T \ln^{\beta_0}\left(\Delta t\right)
+\,\bar{r}_T \ln^{\beta_0-1}\left(\Delta t\right)
\Big),
 \label{XiJ2T}
\end{eqnarray}
where \begin{eqnarray}&& \hspace{-2em}\bar{r}_T=\frac{\beta_0 k_B T}{\hbar \Delta}\Big(\pi \sin \left(\frac{\pi \alpha_0}{2}\right)\Gamma\left(\alpha_0-2\right)-\,2\cos \left(\frac{\pi \alpha_0}{2}\right)\Gamma^{(1)}\left(\alpha_0-2\right)\Big). \nonumber
\end{eqnarray}
If the Ohmicity parameter $\alpha_0$ is not an odd natural number, the dominant part of the above relaxation is equivalent to the following one, $\Xi_T(t)\sim 
\Xi_T\left(\infty\right)+ w_0 r_T\left(\Delta t\right)^{2-\alpha_0} \ln^{\beta_0}\left(\Delta t\right)$. If the Ohmicity parameter $\alpha_0$ takes odd natural values, Eq. (\ref{XiJ2T}) gives $\Xi_0(t)\sim 
\Xi_0\left(\infty\right)+w_0\bar{r}_1\left(\Delta t\right)^{2-\alpha_0} 
\ln^{\beta_0-1}\left(\Delta t\right)$, and provides the power laws
$\Xi_0(t)\sim 
\Xi_0\left(\infty\right)+w_0\bar{r}_1\left(\Delta t\right)^{2-\alpha_0}$ for $\beta_0=1$. Notice the expected similarities between the Eqs. (\ref{Xi004T}) and (\ref{XiJ2T}).

Numerical computations of the coherence term $\rho_{1,0}(t)$ are plotted in Figure \ref{Fig1}. Numerical analysis of the dephasing factor are displayed In Figures \ref{Fig2} and \ref{Fig3}. The short time quadratic growth is confirmed by the parallel asymptotic lines appearing in Figure \ref{Fig2}. The long time logarithmic relaxations result in the asymptotic lines plotted in  Figure \ref{Fig3}.

\section{Regular patterns in the long time information flow}\label{4}

For the system under study the trace distance measure of non-Markovianity that is defined in Refs. \cite{BreuernnMarkovPRL2009,ManiscalcoNJP2015} takes a simple expression in terms of the dephasing rate and dephasing factor and the non-Markovianity measure results in the following form \cite{nnMarkovNeg2LZouPRA2011,FranchiniPRA2013,ManiscalcoPRAr2013},
\begin{equation}
\mathcal{N}=\int_{\gamma(t)<0}\left|\gamma(t)\right|e^{-\Xi(t)} d t. \label{N}
\end{equation}
The open dynamics is Markovian if the dephasing rate is non-negative. 
On the contrary, persistent negative values of the dephasing rate are source of non-Markovianity and are interpreted as a flow of information from the environment back into the system. 
At zero temperature, $T=0$, the open dynamics is Markovian if the function $J\left(\omega\right)/\omega$ is non-increasing. If the SD is differentiable this condition reads as below,
\begin{eqnarray}
J^{\prime}\left(\omega\right)\leq \frac{J\left(\omega\right)}{\omega},
\label{Mcond0}
\end{eqnarray}
for every $\omega>0$. At non-vanishing temperatures, $T>0$, the open dynamics is Markovian if the function $J_T\left(\omega\right)/\omega$ is non-increasing. % for every $\omega>0$. 
If the SD is differentiable, this requirement results in the following constraint,
\begin{eqnarray}
J^{\prime}\left(\omega\right)\leq \left(\frac{1}{\omega}+\frac{\hbar}{k_B T}\operatorname{cosech} \frac{\hbar \omega}{k_BT}\right)J\left(\omega\right), \label{McondT}
\end{eqnarray}
for every $\omega>0$.
Consequently, if the open dynamics is non-Markovian, the function $J\left(\omega\right)/\omega$, for $T=0$, or the function $J_T\left(\omega\right)/\omega$, for $T>0$, is increasing in an interval of frequencies, at least. Let the SD be differentiable for every $\omega>0$. If the open dynamics is non-Markovian the constraint (\ref{Mcond0}), for $T=0$, or (\ref{McondT}), for $T>0$, is not fulfilled for one value of the frequency, at least.

In general, the asymptotic behavior of the dephasing rate depends on integral properties of the SDs. Over long times, $t \gg 1/\Delta$,
the dephasing rate vanishes at zero temperature, $T=0$, if the following condition is fulfilled,
\begin{eqnarray}
\int_0^{\infty}\frac{ J\left(\omega\right)}{\omega}\,
d \omega<\infty.\label{cond3}
\end{eqnarray}
Same behavior is obtained at non-vanishing temperature, $T>0$, if
\begin{eqnarray}
\int_0^{\infty}\frac{ J_T\left(\omega\right)}{\omega}\,
d \omega<\infty.\label{cond3T}
\end{eqnarray}
The long time relaxations depend on the low frequency structure of the SD. We start the analysis of the decays by considering a reservoir of frequency modes at zero temperature, $T=0$, as external environment, and the first class of SDs introduced in Section \ref{31}. Over short times, $t \ll 1/\Delta$, the dephasing rate increases linearly,	
	\begin{eqnarray}
&&\hspace{-2em}\gamma_0(t)\sim l_0 t.
 \label{gamma0s}
\end{eqnarray}
This behavior is independent of the low or high frequency structure of the SDs under study.
Over long times, $t \gg 1/ \Delta$, different forms of relaxations are obtained in dependence on the low frequency structure of the SD, given by Eq. (\ref{o0log}).
 If $\alpha_0=n_0=0$ the dephasing rate tends to the following non-vanishing asymptotic value for $t \gg 1/ \Delta$,
\begin{eqnarray}
\hspace{-0.9em}\gamma_0(t)
\sim \frac{\pi c_{0,0} \Delta}{2}. 
 \label{gamma0l1}
\end{eqnarray}
If $\alpha_0>0$ and $\alpha_0$ is not an even natural number, the dephasing rate vanishes for $t\gg1/\Delta$ according to the relaxations as below,
\begin{eqnarray}
\hspace{-0.9em}\gamma_0(t)\sim c_{0,n_0} g_1 \,\left(\Delta t\right)^{-\alpha_0}
\ln^{n_0}\left(\Delta t\right), 
 \label{gamma0l2}
\end{eqnarray}
that become inverse power laws for $n_0=0$,
\begin{eqnarray}
\hspace{-0.9em}\gamma_0(t)\sim c_{0,n_0} g_1 \,\left(\Delta t\right)^{-\alpha_0}, 
 \label{gamma0l2pl}
\end{eqnarray}
where $g_1=\Delta \sin\left(\pi \alpha_0/2\right)\Gamma\left(\alpha_0\right) $. 
Notice that Eq. (\ref{gamma0l1}) is recovered from Eq. (\ref{gamma0l2pl}) as $\alpha_0\to0^+$. If $\alpha_0=2 m_2$ where $m_2$ and $n_0$ are non-vanishing natural numbers, the dephasing rate vanishes for $t\gg1/\Delta$ as follows,
\begin{eqnarray}
\hspace{-0.9em}\gamma_0(t)\sim c_{0,n_0} g^{\prime}_1 \,\left(\Delta t\right)^{-2 m_2}
\ln^{n_0-1}\left(\Delta t\right), 
 \label{gamma0m2even}
\end{eqnarray}
where $g^{\prime}_1=\pi (-1)^{m_2+1}  n_0\left(2 m_2-1\right)!\Delta/2$.
The above relaxation becomes an inverse power laws for $n_0=1$,
\begin{eqnarray}
\hspace{-0.9em}\gamma_0(t)\sim c_{0,n_0} g^{\prime}_1 \,\left(\Delta t\right)^{-2 m_2}. 
 \label{gamma0m2evenPL}
\end{eqnarray}
If $\alpha_0$ is an even natural number and $n_0$ vanishes, consider the least non-vanishing index $k_2$ such that either $\alpha_{k_2}$ does not take even natural values or 
$\alpha_{k_2}=2m_{k_2}$, where the natural numbers 
$m_{k_2}$ and $n_{k_2}$ do not vanishes. The function $\gamma_{0}(t)$ is obtained in the former case from Eqs. (\ref{gamma0l2}) and
 (\ref{gamma0l2pl}) by substituting the power $\alpha_0$ with $\alpha_{k_2}$ and $n_0$ with $n_{k_2}$, or in the latter case from Eqs. (\ref{gamma0m2even}) and (\ref{gamma0m2evenPL}) by substituting the power $m_2$ with $m_{k_2}$ and $n_0$ with $n_{k_2}$. We consider SDs such that the index $k_2$ exists.

At this stage we focus on SDs that belong to the second class introduced in Section \ref{31}. At zero temperature, $T=0$, we find various forms of relaxations of the dephasing rate for $t \gg 1/\Delta$,
\begin{eqnarray}
\hspace{-1em}\gamma_0(t)\sim 
\frac{w_0}{\left(\Delta t\right)^{\alpha_0}} \Big(g_1 \,\ln^{\beta_0}\left(\Delta t\right)-\bar{g}_1 \,\ln^{\beta_0-1}\left(\Delta t\right)\Big), 
 \label{gamma0l3}
\end{eqnarray}
where $$\bar{g}_1=\beta_0\Delta\left(\frac{\pi}{2} \cos\left(\frac{\pi \alpha_0}{2}\right)\Gamma\left(\alpha_0\right)+\sin\left(\frac{\pi \alpha_0}{2}\right)\Gamma^{(1)}\left(\alpha_0\right)\right).$$ 
If the Ohmicity parameter $\alpha_0$ does not take even natural values, the dominant part of the above asymptotic form is $\gamma_0(t)\sim 
w_0 g_1\left(\Delta t\right)^{-\alpha_0}  \,\ln^{\beta_0}\left(\Delta t\right)$, and becomes the power law $\gamma_0(t)\sim 
w_0 g_1\left(\Delta t\right)^{-\alpha_0}$ if $\beta_0=0$. If the Ohmicity parameter $\alpha_0$ is an even natural number, Eq. (\ref{gamma0l3}) gives $\gamma_0(t)\sim 
-w_0\bar{g}_1\left(\Delta t\right)^{-\alpha_0} \,\ln^{\beta_0-1}\left(\Delta t\right)$ and becomes the power law $\gamma_0(t)\sim 
-w_0\bar{g}_1\left(\Delta t\right)^{-\alpha_0}$ if $\beta_0=1$. Notice the expected similarities between Eqs. (\ref{gamma0l2}) and (\ref{gamma0l3}).

According to the above analysis, at zero temperature, $T=0$, for the first class of SDs under study the information is lost into the environment over short times, $t\ll 1/\Delta$. Over long times, $t\gg 1/\Delta$, the information flows back into the system for the following values of the Ohmicity parameter, $2+4n<\alpha_0<4+4n$, where $n=0,1,2,\ldots$. For the second class of SDs under study we observe the same dependence of the long time information back-flow on the Ohmicity parameter. The corresponding long time dynamics is non-Markovian. The modulus of the coherence term increases, along with the back-flow of information, up to the non-vanishing asymptotic value and recoherence is observed over long times for $2+4n<\alpha_0<4+4n$, at zero temperature. Otherwise, the long time information is lost into the environment, the long time dynamics is Markovian and the modulus of the coherence term decreases down to the asymptotic value. If compared to the initial condition, coherence is partially lost for $\alpha_0>1$. Coherence is fully lost if $0\leq \alpha_0\leq 1$. Notice that in the whole paper the analysis concerns uniquely the short and long time flow of information. Consequently, the dynamics can still be non-Markovian, due to an intermediate back-flow, even if no information flows from the environment back into the system over long times.

\begin{figure}[t]
\centering
\includegraphics[height=6.25 cm, width=10.25 cm]{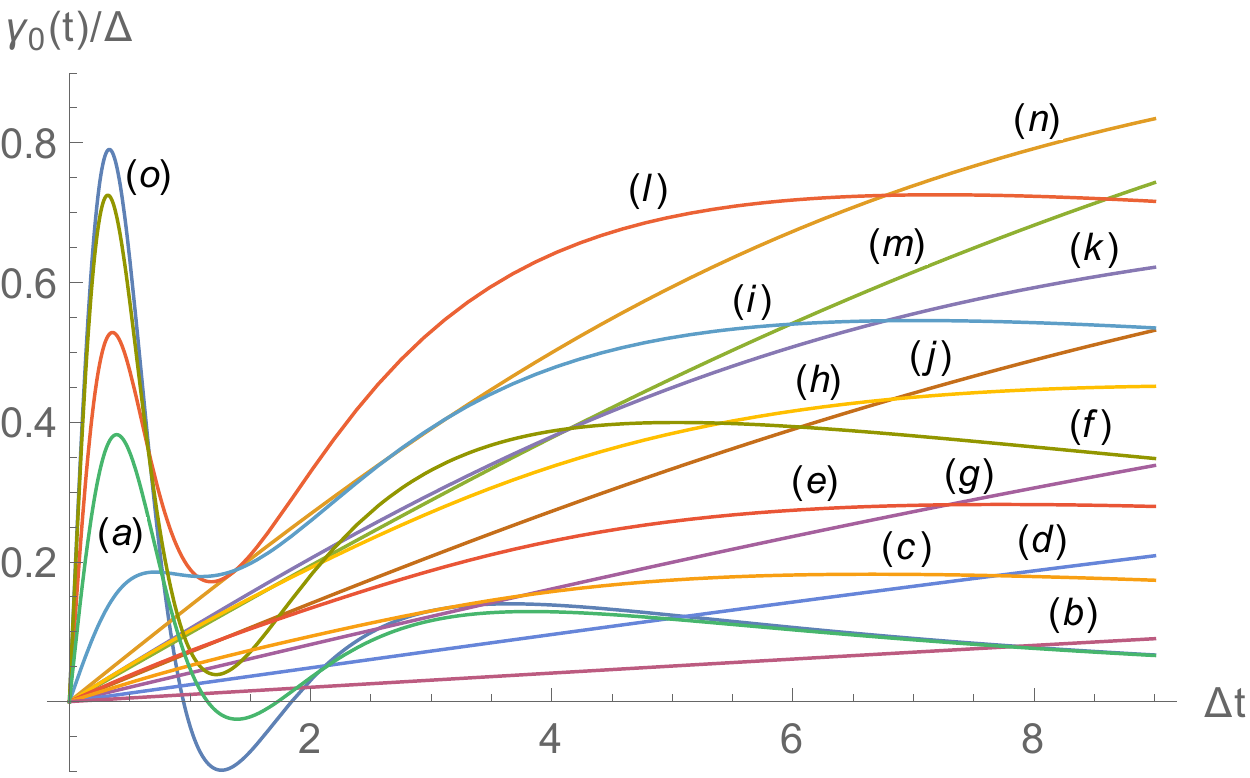}
\vspace*{0cm}
\caption{(Color online) The ratio
$ \gamma_0(t)/\Delta$ versus $\Delta t$ for $0\leq\Delta t\leq 9$, $J\left(\omega\right)=\Delta \left(\omega/\Delta\right)^{\alpha}\exp\left\{- \lambda \omega/\Delta \right\}\ln^2\left(\omega/\Delta \right)$ for different values of the parameters $\alpha$ and $\lambda$. The curve $(a)$ corresponds to $\alpha=2$, $\lambda=1.1$; $(b)$ corresponds to $\alpha=0.9$, $\lambda=40$, $(c)$ corresponds to $\alpha=1.5$, $\lambda=3$; $(d)$ corresponds to $\alpha=0.8$, $\lambda=27$; $(e)$ corresponds to $\alpha=1.3$, $\lambda=2.9$; $(f)$ corresponds to $\alpha=1.3$, $\lambda=0.7$; $(g)$ corresponds to $\alpha=0.8$, $\lambda=17$; $(h)$ corresponds to $\alpha=1.1$, $\lambda=2.9$; $(i)$ corresponds to $\alpha=1.1$, $\lambda=1.2$, $(j)$ corresponds to $\alpha=0.8$, $\lambda=10$; $(k)$ corresponds to $\alpha=0.9$, $\lambda=4.8$; $(l)$ corresponds to $\alpha=1$, $\lambda=0.7$; $(m)$ corresponds to $\alpha=0.7$, $\lambda=9.5$; $(n)$ corresponds to $\alpha=0.8$, $\lambda=4.5$, $(o)$ corresponds to $\alpha=2$, $\lambda=0.9$. }
\label{Fig4}
\end{figure}

\begin{figure}[t]
\centering
\includegraphics[height=6.25 cm, width=10.25 cm]{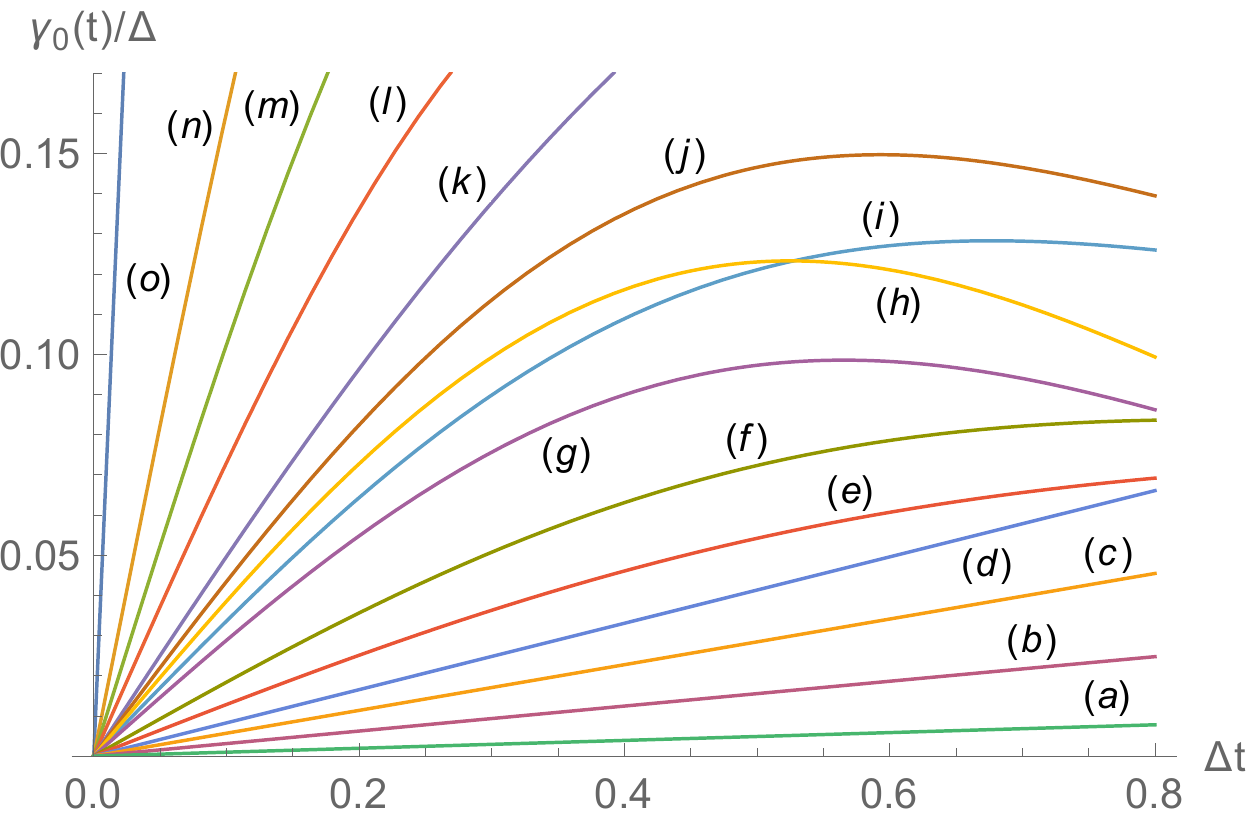}
\vspace*{0cm}
\caption{(Color online) The ratio
$ \gamma_0(t)/\Delta$ versus $\Delta t$ for $0\leq\Delta t\leq 10$, $J\left(\omega\right)=\Delta \left(\omega/\Delta\right)^{\alpha}\exp\left\{- \lambda \omega/\Delta \right\}\ln^2\left(\omega/\Delta \right)$ for different values of the parameters $\alpha$ and $\lambda$. The curve $(a)$ corresponds to $\alpha=1.5$, $\lambda=12$; $(b)$ corresponds to $\alpha=1.5$, $\lambda=5$, $(c)$ corresponds to $\alpha=0.8$, $\lambda=12.5$; $(d)$ corresponds to $\alpha=0.7$, $\lambda=11.5$; $(e)$ corresponds to $\alpha=1.6$, $\lambda=1.9$; $(f)$ corresponds to $\alpha=1.6$, $\lambda=1.7$; $(g)$ corresponds to $\alpha=2$, $\lambda=1.6$; $(h)$ corresponds to $\alpha=2$, $\lambda=1.5$; $(i)$ corresponds to $\alpha=1.4$, $\lambda=1.4$, $(j)$ corresponds to $\alpha=1.4$, $\lambda=1.3$; $(k)$ corresponds to $\alpha=0.8$, $\lambda=1.2$; $(l)$ corresponds to $\alpha=1.3$, $\lambda=1.1$; $(m)$ corresponds to $\alpha=1.3$, $\lambda=1$; $(n)$ corresponds to $\alpha=1$, $\lambda=0.8$, $(o)$ corresponds to $\alpha=1$, $\lambda=0.5$. Over early times each curve tends to an asymptotic line.}
\label{Fig5}
\end{figure}

\begin{figure}[t]
\centering
\includegraphics[height=6.25 cm, width=10.25 cm]{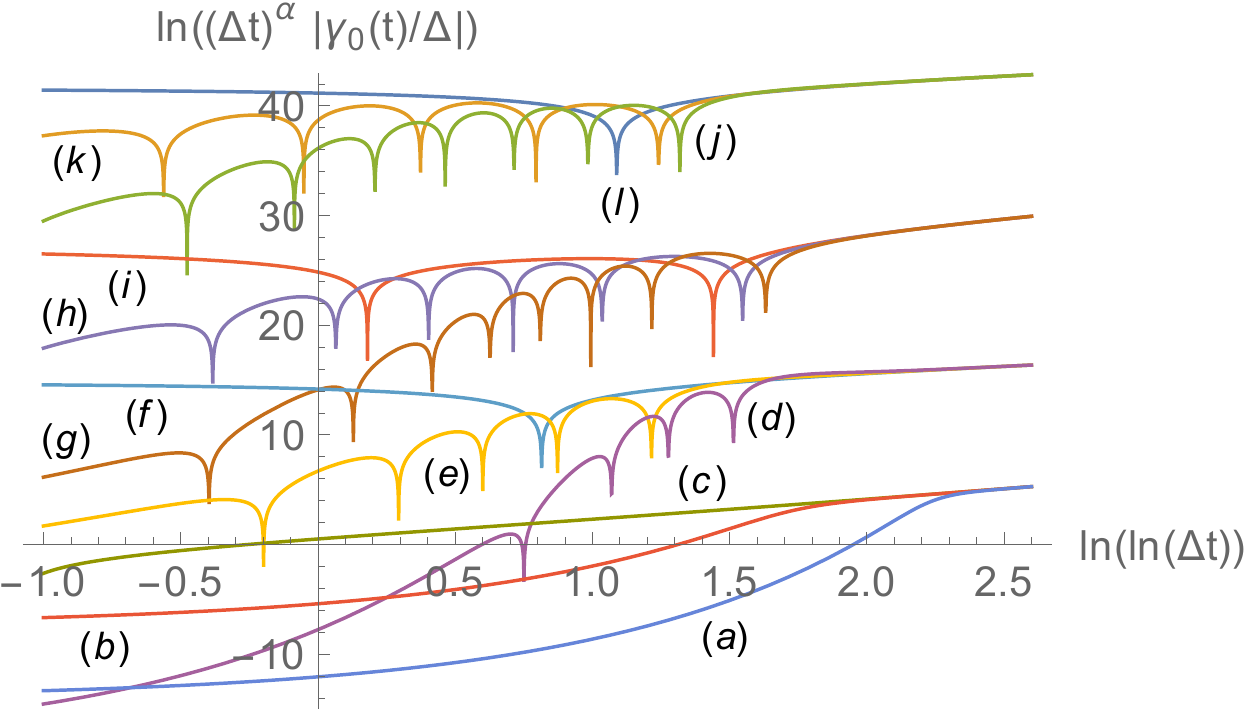}
\vspace*{0cm}
\caption{(Color online) The quantity
$ \ln\left( \left(\Delta t\right)^{\alpha}\gamma_0(t)/\Delta\right)$ versus $\ln \left(\ln\left(\Delta t\right)\right)$ for $\exp\left\{1/e\right\}\leq\Delta t\leq\exp\left\{\exp\left\{2.6\right\}\right\}$, $J\left(\omega\right)=\Delta \left(\omega/\Delta\right)^{\alpha}\exp\left\{- \lambda \omega/\Delta \right\}\ln^2\left(\omega/\Delta \right)$ and different values of the parameters $\alpha$ and $\lambda$. The curve $(a)$ corresponds to $\alpha=1$, $\lambda=10000$; $(b)$ corresponds to $\alpha=1$, $\lambda=200$, $(c)$ corresponds to $\alpha=1$, $\lambda=0.01$; $(d)$ corresponds to $\alpha=10$, $\lambda=20$; $(e)$ corresponds to $\alpha=10$, $\lambda=4$; $(f)$ corresponds to $\alpha=10$, $\lambda=0.01$; $(g)$ corresponds to $\alpha=15$, $\lambda=5$; $(h)$ corresponds to $\alpha=15$, $\lambda=2$; $(i)$ corresponds to $\alpha=15$, $\lambda=0.01$, $(j)$ corresponds to $\alpha=20$, $\lambda=2$; $(k)$ corresponds to $\alpha=20$, $\lambda=1$; $(l)$ corresponds to $\alpha=20$, $\lambda=0.001$. Over long times each curve tends to an asymptotic line.}
\label{Fig6}
\end{figure}

\subsection{Thermal bath}

Let the external environment be a thermal bath, $T>0$. For SDs that belong to the first class under study and $\alpha_0>0$ the dephasing rate increases linearly over short times, $t \ll 1/\Delta$,
	\begin{eqnarray}
&&\hspace{-2em}\gamma_T(t)\sim l_T t.
 \label{gammaTs}
\end{eqnarray}
This behavior is independent of the low or high frequency structure of the SDs under study.
Over long times, the dephasing rate divergences or vanishes in dependence on the low frequency profile of the SD that is given by Eq. (\ref{o0log}).
If $0<\alpha_0<1$ the dephasing rate diverges for $t \gg 1/ \Delta$ according to the following form,
\begin{eqnarray}
\hspace{-0.9em}\gamma_T(t)
\sim c_{0,n_0} g_T \left(\Delta t\right)^{1-\alpha_0}
\ln^{n_0}\left(\Delta t\right), 
 \label{gammaTl1}
\end{eqnarray}
that describes power laws for $n=0$,
\begin{eqnarray}
\hspace{-0.9em}\gamma_T(t)
\sim c_{0,n_0} g_T \left(\Delta t\right)^{1-\alpha_0}. 
 \label{gammaTl1pl}
\end{eqnarray}
The coefficient $g_T$ reads $g_T=2 k_B T\cos\left(\pi \alpha_0/2\right)\Gamma\left(\alpha_0\right)/\left(\hbar\left(1-\alpha_0\right)\right)$. 
If $\alpha_0=1$ the dephasing rate diverges for $t \gg 1/ \Delta$ as below,
\begin{eqnarray}
\hspace{-0.9em}\gamma_T(t)
\sim \frac{c_{0,n_0}\pi k_B T}{\hbar}
\ln^{n_0}\left(\Delta t\right). 
 \label{gammaTl2n}
\end{eqnarray}
If $\alpha_0=1$ and $n_0=0$ the dephasing rate converges for $t \gg 1/ \Delta$ to the following 
non-vanishing value,
\begin{eqnarray}
\hspace{-0.9em}\gamma_T(t)
\sim \frac{c_{0,n_0}\pi k_B T}{\hbar}.
 \label{gammaTl2n0c}
\end{eqnarray}
If $\alpha_0>1$ and $\alpha_0$ is not an odd natural number, the dephasing factor vanishes for $t \gg 1/ \Delta$ according to Eq. (\ref{gammaTl1}).
If $\alpha_0=1+2 m_3$ where $m_3$ and $n_0$
 are non-vanishing natural numbers, the dephasing rate vanishes
 for $t\gg1/\Delta$ as follows,
\begin{eqnarray}
\hspace{-0.9em}\gamma_T(t)\sim c_{0,n_0} g^{\prime}_T \,
\left(\Delta t\right)^{-2 m_3}
\ln^{n_0-1}\left(\Delta t\right), 
 \label{gammaTm3odd}
\end{eqnarray}
where $g^{\prime}_T=\pi (-1)^{1+m_3}k_B T n_0\left(2 m_3-1\right)!/\hbar $.
The above relaxation provides inverse power laws for $n_0=1$,
\begin{eqnarray}
\hspace{-0.9em}\gamma_0(t)\sim c_{0,n_0} g^{\prime}_T \,
\left(\Delta t\right)^{-2 m_2}. 
 \label{gammaTm3oddPL}
\end{eqnarray}
If $\alpha_0$ is an odd natural number and $n_0$ vanishes, consider the least non-vanishing index $k_3$ such that either $\alpha_{k_3}$ does not take odd natural values or $\alpha_{k_3}=1+2m_{k_3}$, where the natural numbers $m_{k_3}$ and $n_{k_3}$ do not vanishes. The function
$\gamma_{T}(t)$ is obtained in the former case from Eqs. (\ref{gammaTl1}) and (\ref{gammaTl1pl}) by substituting the power $\alpha_0$ with $\alpha_{k_3}$ and $n_0$ with $n_{k_3}$, or in the latter
 case from Eqs. (\ref{gammaTm3odd}) and (\ref{gammaTm3oddPL}) by substituting the power $m_3$ with $m_{k_3}$ and $n_0$ with
 $n_{k_3}$. We consider SDs such that the index $k_3$ exists.

Let the external environment consist in a thermal bath, $T>0$, and the auxiliary functions $\Omega_T\left(\nu\right)$ belong to the second class under study. The dephasing rate vanishes for $t \gg 1/\Delta$ according to arbitrary powers of logarithmic forms,
\begin{eqnarray}
\gamma_T\left(t\right)\sim w_0 \left(\Delta t\right)^{1-\alpha_0} \left(g_T \,\ln^{\beta_0}\left(\Delta t\right)
+ \bar{g}_T \,\ln^{\beta_0-1}\left(\Delta t\right)\right),
 \label{gammaTasymptL1}
\end{eqnarray}
where $$\bar{g}_T=\frac{k_B T \beta_0}{\hbar}\, \left(
2 \cos \left(\frac{\pi \alpha_0}{2}\right)\Gamma^{(1)}\left(\alpha_0-1\right)-\pi \sin \left(\frac{\pi \alpha_0}{2}\right)\Gamma\left(\alpha_0-1\right)\right).$$
If the Ohmicity parameter $\alpha_0$ does not take odd natural values, the dominant part of the above relaxation is $\gamma_T\left(t\right)\sim w_0 g_T\left(\Delta t\right)^{1-\alpha_0} \ln^{\beta_0}\left(\Delta t\right)$, and becomes the power law $\gamma_T\left(t\right)\sim w_0 g_T\left(\Delta t\right)^{1-\alpha_0}$ if $\beta_0=0$. If the Ohmicity parameter $\alpha_0$ is an odd natural number, Eq. (\ref{gammaTasymptL1}) gives $\gamma_T\left(t\right)\sim w_0 \bar{g}_T \left(\Delta t\right)^{1-\alpha_0}  \,\ln^{\beta_0-1}\left(\Delta t\right)$ and becomes the power law $\gamma_T\left(t\right)\sim w_0 \bar{g}_T \left(\Delta t\right)^{1-\alpha_0}$ if $\beta_0=1$. Notice the expected similarities between Eqs. (\ref{gammaTl1}) and (\ref{gammaTasymptL1}).

The above results show that at non-vanishing temperatures, $T>0$, and for the first class of SDs under study, the information flows into the environment over short times, $t\ll 1/\Delta$. Over long times, $t\gg 1/\Delta$, back-flow of information appears for the following values of the Ohmicity parameter, $3+4n<\alpha_0<5+4n$, where $n=0,1,2,\ldots$. Same conditions hold for the appearance of the long time back-flow of information by considering the second class of SDs under study. The corresponding long time evolution is non-Markovian. Along with the back-flow of information, the modulus of the coherence term increases up to the non-vanishing asymptotic value and recoherence is observed over long times for $3+4n<\alpha_0<5+4n$, at non-vanishing temperatures. Otherwise, the long time information is lost into the environment, the long time dynamics is Markovian and the modulus of the coherence term decreases down to the asymptotic value. If compared to the initial condition, coherence is partially lost if $\alpha_0>2$. Coherence is fully lost if $0<\alpha_0\leq 2$. Consider the transition from vanishing to an arbitrary non-vanishing temperature. For the SDs under study we observe that the back-flow of information does not change for $3+4n<\alpha_0<4+4n$, while it is inverted for $2+4n<\alpha_0<3+4n$ and $4+4n<\alpha_0<5+4n$. In the transition from vanishing to an arbitrary non-vanishing temperature, the long time recoherence results to be unaffected in the former condition and it is destroyed in the latter ones.

Numerical computations of the dephasing rate are displayed in Figures \ref{Fig4}, \ref{Fig5} and \ref{Fig6}. The short time linear growth is shown in Figure \ref{Fig5}. The long time logarithmic relaxations are confirmed by the asymptotic lines
 plotted in Figure \ref{Fig6}.

\section{Conclusions}\label{5}

We have considered the local dephasing process of a qubit that interacts with a structured reservoir of frequency modes or a thermal bath. We have studied the coherence between the two energy eigenstates of the qubit and the flow of quantum information by analyzing the dephasing factor and dephasing rate over short and long times. The SDs under study are Ohmic-like, at low frequencies, with additional logarithmic factors that are represented by arbitrarily positive or negative powers of logarithmic forms. In this way, the SDs are approximately proportional to the form %$J\left(\omega\right) \approxprop \Delta \left(\omega/\Delta\right)^{\alpha_0}\left(- \ln \left(\omega/\Delta\right)\right)^{\beta_0}$ 
$\Delta \left(\omega/\Delta\right)^{\alpha_0}\left(- \ln \left(\omega/\Delta\right)\right)^{\beta_0}$ for $\omega \ll \Delta$. The corresponding singularities are removable and provide legitimate SDs that contain, enhance and reduce the low frequency power law profiles of the physically feasible Ohmic-like condition. The SDs are arbitrarily tailored at higher frequencies. 

In general, the full loss or persistence of %partial 
coherence, over long times, is determined by integral and low frequency properties of the SD. Over short times, for the SDs under study, the dephasing factor increases quadratically and the dephasing rate grows linearly in time both at zero and at an arbitrary non-vanishing temperature. Over long times, the evolution of the dephasing factor and dephasing rate exhibits various behaviors that are described by logarithmic and power laws, in dependence on the low frequency structure of the SD and on the temperature of the thermal bath. The information flows into the environment over short times both at vanishing and non-vanishing temperature. Over long times, we have found that regular patterns appear in the direction of the flow of information, back into the system or forth into the environment, in dependence on the Ohmicity parameter $\alpha_0$ of the SD, regardless of the logarithmic form factors. At zero temperature, the long time information flows from the environment back into the system in correspondence of the following periodical intervals, $2+4n<\alpha_0<4+4n$, for every $n=0,1,2,\ldots$. At non-vanishing temperatures back-flow of information is obtained over the periodical intervals $3+4n<\alpha_0<5+4n$. In the transition from vanishing to an arbitrary non-vanishing temperature, the back-flow of information stably persists over the intervals $3+4n<\alpha_0<4+4n$. Instead, the back-flow is inverted over the intervals $2+4n<\alpha_0<3+4n$ and $4+4n<\alpha_0<5+4n$. Non-Markovianity and recoherence of the qubit appear along with the back-flow of information. Consequently, recoherence is observed over long times for $2+4n<\alpha_0<4+4n$, at zero temperature, and for $3+4n<\alpha_0<5+4n$, at non-vanishing temperature. For $3+4n<\alpha_0<4+4n$ the transition from vanishing to an arbitrary non-vanishing temperature does not destroy the recoherence process.

The presentation of an experimental setting is beyond the purposes of this paper. Still, the reported results apply to the Ohmic-like SDs of trapped impurity atoms that are immersed in a Bose-Einstein condensate environment. Furthermore, if the low frequency power law profiles of the Ohmic-like SDs are enhanced or reduced via arbitrary positive or negative powers of logarithmic form factors, the direction of the information flow is not altered by the logarithmic terms and depends uniquely on the Ohmicity parameter of the Ohmic-like term. Consequently, the patterns in the information flow remain stable with respect to the mentioned logarithmic perturbations of the Ohmic-like SDs. We believe that the present analysis provides further scenarios for the implementation of a stable control of the flow of quantum information and the appearance of non-Markovian dynamics and recoherence via the engineering reservoir approach.

\vspace{0 mm}

\appendix\label{A}
\section{details}
The evolution of the reduced density matrix $\rho(t)$ is given by the master equation (\ref{Eq1}). The off-diagonal elements of the reduced density matrix are described by Eq. (\ref{rho01t}) in terms of the dephasing factor $\Xi(t)$. This function is given by Eq. (\ref{Xi0t}), for $T=0$, and Eq. (\ref{XiTt}), for $T>0$. If the second negative moment of the SD is finite, Eq. (\ref{cond1}), the expression $\int_0^{\infty} J\left(\omega\right)\cos\left(\omega t \right)/\omega^2 d\omega$ 
vanishes over long times due to the Riemann-Lebesgue lemma. In this way, Eq. (\ref{rhoinfty}) is obtained. Again, according to the Riemann-Lebesgue lemma, if the second negative moment of the effective SD is finite, Eq. (\ref{cond1T}), the expression $\int_0^{\infty} J_T\left(\omega\right)\cos\left(\omega t \right)/\omega^2 d\omega$ vanishes over long times. In this way, Eq. (\ref{rhoinftyT}) is obtained.

The asymptotic behavior of the function $\Xi_0(t)$ is studied in the dimensionless variables $\nu=\omega/\Delta$ and $\tau=\Delta t$ by considering the function $F_0\left(\tau\right)$, that is defined as $F_0\left(\tau\right)=\Xi_0\left(\tau/\Delta \right)$. According to this definition, the function reads
\begin{eqnarray}
\hspace{-1em}F_0\left(\tau\right)=  \int_0^{\infty}
\frac{\Omega\left(\nu\right)}{\nu^2}\sin^2 
\left(\frac{	\tau \nu}{2}\right) \, d\nu.\label{MXit}
\end{eqnarray}
 The Mellin transform \cite{BleisteinBook,Wong-BOOK1989} of the function $F_0\left(\tau\right)$ is defined as follows, $\hat{F}_0\left(s\right)= \int_0^{\infty}\tau^{s-1}F_0\left(\tau\right) \, d \tau$, and reads
\begin{eqnarray}
\hspace{-1em}\hat{F}_0(s)= -\cos\left(\frac{\pi s}{2}\right) \,\Gamma\left(s\right) \hat{\Omega}\left(-1-s\right).  \label{MXis}
\end{eqnarray}
The fundamental strip depends on the asymptotic behavior of the auxiliary function \cite{BleisteinBook,Wong-BOOK1989}. Consider the first class of SDs introduced in Section \ref{31} via the asymptotic form (\ref{o0log}). The fundamental strip of the Mellin transform $\hat{F}_0(s)$ is $\min\left\{0,\alpha_0-1\right\}>\operatorname{Re}s >-2$. The following asymptotic relationship \cite{GradRyz},
 \begin{eqnarray}
&&\hspace{-4em}\left|\cos\left(\frac{\pi s}{2}\right) \,\Gamma\left(s\right)\right| \sim \left|\sin\left(\frac{\pi s}{2}\right) \,\Gamma\left(s\right)\right| %\nonumber \\ &&\hspace{3.4em}
\sim \left(\frac{\pi}{2}\right)^{1/2}\left|\operatorname{Im} s\right|^{\operatorname{Re}s-1/2}, \label{SimEq1}
\end{eqnarray}
 holds for $\left|\operatorname{Im} s\right|\to +\infty$. For $\max\left\{-4,-2-\chi_0\right\}<\operatorname{Re} s<\min\left\{-1/2-\epsilon_0,\alpha_0-1\right\} $ and $\left|\operatorname{Im} s	\right|\to+\infty$ the Mellin transform of the function $F_0(t)$ vanishes as follows, $\hat{F}_0\left(s\right)= o\left(\left|\operatorname{Im} s\right|^{-1-\epsilon_0}\right)$, where $\epsilon_0\in \left(0,3/2\right)$. Consequently, the function $\hat{F}_0\left(s\right)$ decreases sufficiently fast in the strip as $|\operatorname{Im} s|\to+\infty$ and the singularity in $s=-2$ provides the asymptotic expansion of the dephasing factor at short times, given by Eq. (\ref{Xi000}). As far as the long time evolution is concerned, let the strip $ \mu_0\leq \operatorname{Re} s\leq\delta_0$ exist such that the function $\hat{\Omega}\left(-1-s\right)$, or the meromorphic continuation, vanishes as follows,
\begin{eqnarray}
\hspace{-1em}\hat{\Omega}\left(-1-s\right)= O\left(\left|\operatorname{Im} s\right|^{-\zeta_0}\right), \label{c1Omegas}
\end{eqnarray}
 for $|\operatorname{Im} s|\to+\infty$, where $\zeta_0>1/2+\delta_0$. The parameters $\mu_0$ and $\delta_0$ fulfill the constraints as below, $\mu_0 \in \left(-2,\min\left\{0,\alpha_0-1\right\}\right)$, $\delta_0 \in \left(\alpha_0-1,0\right)$ for $0\leq \alpha_0<1$, or $ \delta_0\in \left(\alpha_{k_4},\alpha_{k_5}\right)$ if $\alpha_0\geq 1$. The parameter $\alpha_{k_4}$ coincides with the positive power $\alpha_0$ if  $\alpha_0$ is not an even natural number, or if $\alpha_0=2m_0$ and $n_0>0$, otherwise $\alpha_{k_4}$ coincides with the parameter $\alpha_{k_0}$ that is defined in Section \ref{3}.  The index $k_4$ 
is the least natural number that is larger than $k_3$ and such 
that $\alpha_{k_4}$ is not an even natural number, or such 
that $\alpha_{k_4}$ is an even natural 
number and $n_{k_4}> 0$. Under the above conditions, the singularity of the function $\hat{F}_0(s)$ in $s=\alpha_0-1$ and $0\leq \alpha_0\leq 1$, in $s=0$ and $s=\alpha_{k_4}-1$ if $\alpha_0>1$, provides the asymptotic forms given by Eqs. (\ref{Xi00})-(\ref{Xi0evenPL}).

At non-vanishing temperature the asymptotic behavior of the dephasing factor is evaluated via the function $F_T\left(\tau\right)$, defined as $F_T\left(\tau\right)=\Xi_T\left(\tau/\Delta \right)$, and the Mellin transform, $\hat{F}_T(s)$, that reads
\begin{eqnarray}
\hspace{-1em}\hat{F}_T(s)= -\cos\left(\frac{\pi s}{2}\right) \,\Gamma\left(s\right) \hat{\Omega}_T\left(-1-s\right).  \label{MXis}
\end{eqnarray}
The fundamental strip is $\min\left\{0,\alpha_0-2\right\}>\operatorname{Re}s >-2$, for $\alpha_0>0$. %, or $0>\operatorname{Re}s >-2$ if $\lambda>0$.
The relationship (\ref{SimEq1}) implies that for $\max\left\{-4,-2-\chi_0\right\}<\operatorname{Re} s<\min\left\{-1/2-\epsilon_1,\alpha_0-2\right\} $ and $|\operatorname{Im} s|\to+\infty$ the Mellin transform of the function $F_T(t)$ vanishes as follows, $\hat{F}_T\left(s\right)= o\left(\left|\operatorname{Im} s\right|^{-1-\epsilon_1}\right)$, where $\epsilon_1\in \left(0,3/2\right)$. Consequently, the function $\hat{F}_T\left(s\right)$ decreases sufficiently fast in the strip as $|\operatorname{Im} s|\to+\infty$ and the singularity in $s=-2$ provides the asymptotic expansion of the dephasing factor at short times, given by Eq. (\ref{Xi000T}). As far as the long time behavior is concerned, let the strip $\mu_1\leq \operatorname{Re} s\leq \delta_1 $ exist such that the function $\hat{\Omega}_T\left(-1-s\right)$, or the meromorphic continuation, vanishes as follows,
\begin{eqnarray}
\hspace{-1em}\hat{\Omega}_T\left(-1-s\right)= O\left(\left|\operatorname{Im} s\right|^{-\zeta_1}\right), \label{c1Omegas}
\end{eqnarray}
 for $|\operatorname{Im} s|\to+\infty$, where $\zeta_1>1/2+\delta_1$. The parameters $\mu_1$ and $\delta_1$ fulfill the following constraints, $\mu_1 \in \left(-2,\alpha_0-2\right)$, $\delta_1 \in \left(\alpha_0-2,0\right)$ for $0<\alpha_0<2$, or $\mu_1 \in \left(-2,0\right)$ and $ \delta_1\in \left(\alpha_{k_6},\alpha_{k_7}\right)$ for $\alpha_0\geq 2$. The parameter $\alpha_{k_6}$
coincides with the positive power $\alpha_0$ if
 $\alpha_0$ is not an odd natural number, or if $\alpha_0=1+2m_1$  and $n_0>0$, otherwise $\alpha_{k_6}$ coincides with the parameter $\alpha_{k_1}$ that is defined in Section \ref{3}. The index $k_7$ is the least natural number that is larger than $k_6$ and such 
that $\alpha_{k_7}$ is not an odd natural number, or such 
that $\alpha_{k_7}$ is an odd natural number and $n_{k_7}> 0$. Under the above conditions, the singularity of the function $\hat{F}_T(s)$ in $s=\alpha_0-2$ for $0<\alpha_0\leq 2$, or in $s=0$ and $s=\alpha_{k_6}-2$ for $\alpha_0>2$ provides Eqs. (\ref{Xi00T})-(\ref{XiToddPL}).

For the second class of SDs introduced in Section \ref{31} the study performed in Refs. \cite{WangLinJMAA1978,Wong-BOOK1989}, allows the asymptotic analysis of the expression (\ref{Xi0t}), for $T=0$, and (\ref{XiTt}), for $T>0$, in terms of the dimensionless variables $\nu$ and $\tau$. In this way, the asymptotic forms (\ref{Xi0J2}), for $T=0$, and (\ref{XiJ2T}), for $T>0$, are obtained.

The dephasing rate $\gamma(t)$ is defined by Eq. (\ref{gamma0}), for $T=0$, and by Eq. (\ref{gammaT}), for $T>0$. The constraints (\ref{Mcond0}) and (\ref{McondT}) are obtained by observing that the sine transforms of non-increasing functions are non-negative. For the first class of SDs under study the asymptotic behavior of the dephasing rate $\gamma_0(t)$ is studied by considering the function $G_0\left(\tau\right)$, that is defined as $G_0\left(\tau\right)=\gamma_0\left(\tau/\Delta\right)$ and reads
\begin{eqnarray}
&&\hspace{-3em}G_0\left(\tau\right)=\Delta\int_0^{\infty}\frac{\Omega\left(\nu\right)}{\nu} \, \sin \left(\nu \tau\right) \,d \nu.\label{gammaO0}
\end{eqnarray}
The Mellin transform $\hat{G}_0\left(s\right)$ results as below,
\begin{eqnarray}
&&\hspace{-3em}\hat{G}_0\left(s\right)=\Delta
\sin\left(\frac{\pi s}{2}\right) \,\Gamma\left(s\right) \hat{\Omega}\left(-s\right).  \label{Gs} 
\end{eqnarray}
The fundamental strip is $\min\left\{1,\alpha_0\right\}>\operatorname{Re}s >-1$. The relationship (\ref{SimEq1}) suggests that for $\max\left\{-3,-1-\chi_0\right\}<\operatorname{Re} s<-1/2-\epsilon_2 $ and $|\operatorname{Im} s|\to+\infty$ the Mellin transform of the function $G_0(t)$ vanishes as follows, $\hat{G}_0\left(s\right)=o\left(\left|\operatorname{Im} s\right|^{-1-\epsilon_2}
\right)$, where $\epsilon_2 \in \left(0,1/2\right)$. Consequently, the function $\hat{G}_0\left(s\right)$  decreases sufficiently fast in the strip as $|\operatorname{Im} s|\to+\infty$ and the singularity in $s=-1$ provides Eq. (\ref{gamma0s}). 
 As far as the long time evolution is concerned, let the strip $\mu_2\leq \operatorname{Re} s\leq \delta_2$ exist,
such that the function $\hat{\Omega}\left(-s\right)$, or the meromorphic continuation, vanishes as follows,
\begin{eqnarray}
\hspace{-1em}\hat{\Omega}\left(-s\right)= O\left(\left|\operatorname{Im} s\right|^{-\zeta_2}\right), \label{c1Omegas}
\end{eqnarray}
 for $|\operatorname{Im} s|\to+\infty$, where $\zeta_2>1/2+\delta_2$. The parameters $\mu_2$ and $\delta_2$ fulfill the constraints as below, $\mu_2 \in \left(-1,\min\left\{1,\alpha_0\right\}\right)$, $ \delta_2\in \left(\alpha_{k_8},\alpha_{k_9}\right)$. The parameter $\alpha_{k_8}$ coincides with the positive power $\alpha_0$ if $\alpha_0$ is not an even natural number, or if $\alpha_0=2m_2$  and $n_0>0$, otherwise $\alpha_{k_8}$ coincides with the power $\alpha_{k_2}$ that is defined in Section \ref{3}. The index $k_9$ is the least natural number that is larger than $k_8$ and such that $\alpha_{k_9}$ is not an even natural number, or such that $\alpha_{k_9}$ is an even natural number and $n_{k_9}> 0$. Under the above condition, the singularity of the function $\hat{G}_0(s)$
 in $s= \alpha_{k_8}$ provides the asymptotic forms given by Eqs. (\ref{gamma0l1})-(\ref{gamma0m2evenPL}).

For non-vanishing temperatures, $T>0$, we study the function $G_T\left(\tau\right)$, that is defined as $G_T\left(\tau\right)=\gamma_T\left(\tau/\Delta\right)$ and reads \begin{eqnarray}
&&\hspace{-4em}G_T\left(\tau\right)=\Delta\int_0^{\infty}\frac{\Omega\left(\nu\right)}{\nu} \, \coth \frac{\hbar \Delta \nu}{2 k_B T}\, \sin \left(\nu \tau\right) \,d \nu. \label{gammaOT}
\end{eqnarray}
The Mellin transform $\hat{G}_T\left(s\right)$ results in the following form,
\begin{eqnarray}
&&\hspace{-3em}\hat{G}_T\left(s\right)=\Delta
\sin\left(\frac{\pi s}{2}\right) \,\Gamma\left(s\right) \hat{\Omega}_T\left(-s\right).  \label{GTs} 
\end{eqnarray}
The fundamental strip is $\min\left\{1,\alpha_0-1\right\}>\operatorname{Re}s >-1$, where $\alpha_0>0$. The relationship (\ref{SimEq1}) implies that for $\max\left\{-3,-1-\chi_0\right\}<\operatorname{Re} s<\min \left \{-1/2-\epsilon_3,\alpha_0-1 \right\}$ and $|\operatorname{Im} s|\to+\infty$ the Mellin transform of the function $G_T(t)$ vanishes as follows, $\hat{G}_0\left(s\right)=o\left(\left|\operatorname{Im} s\right|^{-1-\epsilon_3}
\right)$, where $\epsilon_3 \in \left(0,1/2\right)$. Consequently, the function $\hat{G}_T\left(s\right)$ decreases sufficiently fast in the strip as $|\operatorname{Im} s|\to+\infty$ and the singularity in $s=-1$ gives Eq. (\ref{gammaTs}). As far as the long time behavior is concerned, let the strip $\mu_3\leq \operatorname{Re} s\leq \delta_3$ exist such that the function $\hat{\Omega}\left(-s\right)$, or the meromorphic continuation, vanishes as follows, \begin{eqnarray}
\hspace{-1em}\hat{\Omega}_T\left(-s\right)= O\left(\left|\operatorname{Im} s\right|^{-\zeta_3}\right), \label{c1Omegas}
\end{eqnarray}
 for $|\operatorname{Im} s|\to+\infty$, where $\zeta_3>1/2+\delta_3$. The parameters $\mu_3$ and $\delta_3$ fulfill the following constraints, $\mu_3 \in \left(-1,\min\left\{1,\alpha_0-1\right\}\right)$ 
for $\alpha_0>0$, $ \delta_3\in \left(\alpha_{k_{10}},\alpha_{k_{11}}\right)$. The parameter $\alpha_{k_{10}}$ coincides with the positive power $\alpha_0$ if $\alpha_0$ is not an odd natural number, or if $\alpha_0=1+2m_3$  and $n_0>0$, otherwise $\alpha_{k_{10}}$ coincides with the power $\alpha_{k_3}$ that is defined in Section \ref{3}. The index $k_{11}$ is the least natural number that is larger than $k_{10}$ and such that $\alpha_{k_{11}}$ is not an odd natural number, or such that $\alpha_{k_{11}}$ is an odd natural number and $n_{k_{11}}> 0$.  Under the above conditions the singularity of the function $\hat{G}_T(s)$ in $s=\alpha_{k_{10}}-1$ provides  the asymptotic forms given by Eqs. (\ref{gammaTl1})-(\ref{gammaTm3oddPL}).

For the second class of SDs introduced in Section \ref{31}, the long time behavior of the dephasing rate is obtained from the study performed in Refs. \cite{WangLinJMAA1978,Wong-BOOK1989} in terms of the dimensionless variables $\nu$ and $\tau$. In this way, we obtain the expressions (\ref{gamma0l3}), for $T=0$, and (\ref{gammaTasymptL1}), for $T>0$.

 The direction of the flow of information over short and long times is performed by studying the sign of the first term of the asymptotic expansion over short and long times, respectively. Negatives values correspond to back-flow of information. This concludes the demonstration of the present results.

\end{document}